\documentclass[11pt]{article}

\usepackage[preprint]{acl}

\usepackage{times}
\usepackage{latexsym}

\usepackage{url}
\usepackage{multirow}
\usepackage[table,xcdraw]{xcolor}

\usepackage[T1]{fontenc}
\usepackage[utf8]{inputenc}

\usepackage{microtype}

\usepackage{inconsolata}

\usepackage{graphicx}

\title{When Single Agent Is Enough: Rethinking Multi-Agent Systems for Automated Code Review}

\author{Jiawei Li \\
  University of California, Irvine \\
  \texttt{jiawl28@uci.edu} \\\And
   Hoa Khanh Dam \\
  University of Wollongong \\
  \texttt{hoa@uow.edu.au} \\\And
  Iftekhar Ahmed \\
  University of California, Irvine \\
  \texttt{iftekha@uci.edu}
  \\}

\begin{document}
\maketitle
\begin{abstract}
Code review plays a critical role in identifying software quality issues and ensuring system robustness. Prior work has proposed techniques for review comment generation (RCG) and code refinement (CR), with recent multi-agent (MA) approaches achieving state-of-the-art results by coordinating agents specialized in different quality dimensions. However, MA systems introduce substantial architectural complexity and runtime overhead, and recent evidence suggests they may underperform single-agent (SA) approaches in certain tasks. In this work, we systematically compare MA and SA designs for RCG and CR. Surprisingly, our results show that SA approaches perform comparably to—and sometimes outperform—MA systems, despite their simpler design. These findings challenge the prevailing assumption that MA architectures are the default choice for complex software engineering tasks.
%Code review, a process that aims to identify software quality issues, is essential for ensuring overall software quality and robustness. To alleviate the heavy workload of human reviewers, prior research has proposed techniques to automatically generate review comments (Review Comment Generation, RCG) and refine the submitted code patch based on the review comment (Code Refinement, CR), with Multi-Agent (MA) approaches achieving the state-of-the-art performance. Existing approaches enable multiple agents with expertise in different quality dimensions to collaboratively conduct code review. However, MA-based approaches are plagued by high system complexity and runtime overhead, and can underperform Single-Agent systems in certain tasks. In this study, we investigated the performance gap between these two agentic designs on RCG and CR that presumably benefit the most from MA-based approaches. Surprisingly, our results show that SA-based approaches perform comparably to--and in some cases outperform--MA-based techniques, challenging the longstanding belief of MA-based approaches being the default setting for complex software engineering tasks.   
\end{abstract}

\section{Introduction}
\label{sec:intro}
% code review and code review burden -> why we need code review automation

Code review \cite{bacchelli2013expectations, davila2021systematic} is a process in software development where a reviewer carefully examines the contributed patch by a developer to identify quality issues, such as software defects \cite{mantyla2008types}, vulnerabilities \cite{braz2022less}, and formatting inconsistencies \cite{han2020does}, and proposes potential solutions or suggestions to improve the patch. The complete code review process comprises two main tasks: \textit{Review Comment Generation} (RCG) \cite{li2022auger} and \textit{Code Refinement} (CR) \cite{guo2024exploring}. In RCG, the reviewer composes the identified issue descriptions and potential solutions in natural-language text as the review comment. Then, the patch author reviews the comment to address any issues raised and improve code quality in the CR task. If the refined patch resolves all issues in the review comment, the reviewer will accept it and merge it into the software repository. These two tasks play an essential role in ensuring overall software quality by resolving quality issues before patches are merged. Oftentimes, active and experienced reviewers would be invited to review patches \cite{thongtanunam2017review}. These reviewers can be burdened with numerous pending review requests \cite{bosu2016process, macleod2017code}, and such manual inspection of the code can be time-consuming in large projects \cite{bosu2013impact}. These factors can lead to review delays that negatively impact software quality \cite{baysal2016investigating}.

To reduce the heavy workload of human reviewers, prior research has proposed techniques to automate RCG and CR \cite{tufano2021towards, tufano2022using}. With recent advances in Large Language Models (LLMs), many approaches now leverage LLMs and LLM-based agents, achieving state-of-the-art performance \cite{guo2024exploring, tang2024codeagent, ren2025hydra}. Code review typically evaluates multiple software quality dimensions, including functionality, security, robustness, and performance \cite{mantyla2008types, ren2025hydra}. To address this multi-faceted nature, Multi-Agent (MA) frameworks have emerged as a promising paradigm, enabling the generation of more comprehensive and targeted review comments \cite{tang2024codeagent, ren2025hydra}. For instance, the state-of-the-art MA-based system Hydra-Reviewer \cite{ren2025hydra} employs multiple LLM agents to analyze a patch, where each agent focuses on a specific quality dimension. The comments produced by these agents are subsequently aggregated into a final review summary. Moreover, empirical studies show that patches tend to achieve higher quality when reviewed and discussed by multiple reviewers \cite{bavota2015four, kononenko2015investigating}. This observation further motivates MA-based approaches, where multiple agents with diverse expertise collaboratively perform code review \cite{tang2024codeagent, ren2025hydra}.

% However, problems of MAS; people have started questioning MAS; SAS has shown decent performance too due to LLMs being more and more powerful; -> so the question is: does MAS outperform SAS in code review.

%OLD: However, MA-based approaches tend to introduce significant system complexity and runtime overhead compared to Single-Agent (SA) systems. Since LLMs are becoming increasingly powerful at long-context reasoning and tool calling, the superiority that MA-based approaches have traditionally held may rapidly diminish. Moreover, MA-based approaches can even underperform SA ones due to extensive overthinking \cite{gao2025single}. Thus, we ask the following overarching question:

However, MA-based approaches tend to introduce substantial system complexity and runtime overhead compared to Single-Agent (SA) systems. Since modern LLMs have rapidly improved in long-context reasoning and tool use, they potentially narrow the advantages traditionally attributed to MA-based approaches. In some cases, MA-based approaches may underperform compared to SA-based approaches due to excessive deliberation or coordination overhead among agents \cite{gao2025single}. Thus, we ask the following question:

%OLD

\textit{How do MA-based approaches compare to their SA-based counterparts in code review automation tasks that are presumed to benefit from MA approaches?}

%\hl{Following is new}
%\textit{To what extent do MA-based approaches outperform SA-based approaches in automated code review tasks that are presumed to benefit from MA approaches?}

% experiment design - hydra reviewer and its SA counterpart
%New
To answer this question, we analyze two recent state-of-the-art MA-based code review systems: \textbf{\textit{CodeAgent}} \cite{tang2024codeagent} and \textbf{\textit{Hydra Reviewer}} \cite{ren2025hydra}. However, we exclude \textit{CodeAgent} due to reproducibility issues in its released replication package, which generates a substantial number of irrelevant review comments. Excluding it ensures that our evaluation more accurately reflects the effectiveness of MA-based approaches. For the SA counterpart of \textit{Hydra Reviewer}, we adopt one of its baselines that merges the prompts used by agents targeting different quality dimensions into a single prompt provided to the LLM. In addition, since the ReAct paradigm \cite{yao2022react} is widely used to build LLM-based agents, we implement an SA-based ReAct agent whose actions invoke tools that detect multiple quality dimensions and retrieve contextual information to support code comprehension. To construct the MA counterpart, each ReAct agent is equipped with a tool specialized for detecting a single quality dimension, along with tools for retrieving contextual information. This design allows each agent to act as an expert for a specific dimension, similar to \textit{Hydra Reviewer}. Agents operate under a decentralized coordination scheme, where each agent decides whether to transfer control to another agent or generate the final output. Such decentralized MA architectures have recently gained traction \cite{yang2025agentnet,dang2025multi}, as they mitigate scalability limitations and single points of failure associated with centralized controllers \cite{yang2025agentnet}.

With both MA-based approaches and their SA counterparts implemented, we evaluate their effectiveness on RCG and CR. For RCG, we employ multiple LLM-as-a-judge evaluators from prior work, each assessing a specific aspect of review comment quality. These evaluators have been shown to align closely with human expert judgments. For CR, we construct an LLM-as-a-judge by adapting the agents from \cite{ren2025hydra} to detect code quality issues across 18 dimensions commonly considered in human code review. We further incorporate code similarity metrics that compare generated patches with human-authored patches in the dataset. Our experimental results show that SA-based approaches consistently perform comparably to or outperform their MA counterparts on both tasks, despite the latter requiring more complex system architectures and higher computational costs.

\section{MA and SA for Code Review}
\label{sec:methodology}

This section describes the design of the MA-based code review automation techniques evaluated in our experiments and their SA counterparts for both RCG and CR tasks. In RCG, the goal is to generate review comments that identify code quality issues and suggest potential fixes for a given patch. In CR, the objective is to produce a refined patch that addresses the issues described in the review comments. We exclude \textit{CodeAgent} \cite{tang2024codeagent} from our experiments because, based on our manual inspection, its outputs frequently contain comments that are irrelevant to the target patches. Examples of such irrelevant comments are provided in Appendix~\ref{sec:appendix-codeagent}.

%This section details the design of the MA-based code review automation techniques in our experiment and their SA-based counterparts for both RCG and CR tasks. The automation of RCG requires generating a review comment that identifies quality issues and proposes potential solutions for a code patch, while CR aims to generate a refined patch that addresses all issues mentioned in the review comment. It's worth noting that we excluded \textit{CodeAgent} \cite{tang2024codeagent} because, after our manual examination, its generation is often dominated by comments irrelevant to patches. We contacted the authors but received no response. Examples about such comments are provided in the Appendix \ref{sec:appendix-codeagent}.

\subsection{Hydra Reviewer}

\textit{Hydra Reviewer} \cite{ren2025hydra} is a state-of-the-art approach for RCG. The authors construct a taxonomy of 18 code quality dimensions commonly considered in GitHub code reviews. The system adopts an MA architecture, where each LLM-based agent specializes in one dimension and generates a corresponding review comment. The comments are then filtered and summarized into a final comprehensive review. Each agent consists of three prompted LLM modules and receives the patch along with the relevant software context. The generation module produces a comment for the assigned dimension, the reflection module verifies its alignment with that dimension, and the clean-up module removes irrelevant content to produce the final comment. After collecting comments from all agents, an LLM filters and summarizes them into a single review comment. We include brief descriptions of the 18 dimensions in Appendix \ref{sec:appendix-prompts}, but more details can be found in \cite{ren2025hydra}.

%OLD:\textit{Hydra Reviewer} \cite{ren2025hydra} is the most state-of-the-art code review automation approach for RCG. The authors first constructed a taxonomy of 18 quality dimensions targeted by code reviewers in popular GitHub repositories. It's an MA-based approach in which each LLM-based agent focuses on one of 18 quality dimensions to generate a review comment. Then, all 18 review comments are filtered and summarized into a comprehensive code review comment. Specifically, each LLM-based agent consists of three prompted LLMs as functional units and is input with the patch along with the relevant software context. The generation unit is prompted to write a review comment on a patch, focusing only on one designated dimension. The reflection unit prevents the generated comment from deviating from the designated dimension. The generation unit may generate a comment multiple times until the reflection unit approves it. Finally, the clean-up unit removes unhelpful content and produces the final review comment. After obtaining 18 review comments from all agents, the LLM is prompted to select the correct comments and summarize them into a comprehensive review comment. More design details and prompts of \textit{Hydra Reviewer} can be found in \cite{ren2025hydra}.

\noindent\textbf{RCG.} We use the \textit{Hydra Reviewer} implementation from the authors’ replication package to generate review comments for code patches. For the SA counterpart, we adopt \textit{C-ChatGPT}, a baseline provided by the authors. Unlike Hydra Reviewer’s multi-agent design, \textit{C-ChatGPT} uses a single LLM-based agent, where the prompts from the 18 dimension-specific agents are merged into a single prompt to generate review comments covering all quality dimensions. Further details on the design and prompts of \textit{Hydra Reviewer} and \textit{C-ChatGPT} can be found in \cite{ren2025hydra}.

%We took the \textit{Hydra Reviewer} from the replication package published by the authors to generate review comments for code patches. 

%For its SA-based counterpart, we selected an approach variant (\textit{C-ChatGPT}) implemented by the authors as a baseline that only involves a single LLM-based agent, where the prompts used by the 18 agents of \textit{Hydra Reviewer} are combined into a single prompt and fed to the LLM to generate a review comment covering all dimensions. For more information regarding the design and prompts of \textit{Hydra Reviewer} and \textit{C-ChatGPT}, refer to \cite{ren2025hydra}.

\noindent\textbf{CR.} First, we examine the impact of review comments generated by \textit{Hydra Reviewer} and \textit{C-ChatGPT} on a state-of-the-art CR technique named \textit{ChatGPT-CR} \cite{guo2024exploring}. We use the implementation and prompts from the released replication package. Given a patch and its associated review comment, \textit{ChatGPT-CR} prompts ChatGPT to produce a refined patch that addresses the identified issues. By evaluating the quality of the refined patches generated from comments produced by the two RCG techniques, we assess the effectiveness of \textit{Hydra Reviewer} and \textit{C-ChatGPT} for the CR task.

%OLD: First, we investigated the impact of the comments generated by \textit{Hydra Reviewer} and \textit{C-ChatGPT} on a state-of-the-art CR technique, \textit{ChatGPT-CR} \cite{guo2024exploring}. We took the implementation and prompts from its replication package. It takes the patch and the generated review comment as inputs and outputs a refined patch that addresses the issues identified in the review comment, which is achieved by prompting ChatGPT. Analyzing the quality of the refined patch based on the comments from the two RCG techniques would reveal the effectiveness of the \textit{Hydra Reviewer} and \textit{C-ChatGPT} on the task of CR. 

We also evaluate the ability of \textit{Hydra Reviewer} and \textit{C-ChatGPT} to perform CR directly when provided with human-written review comments. To do so, we adapt their prompts to first identify the quality dimensions referenced in the comments and then refine the patch accordingly.

For \textit{Hydra Reviewer}, we modify each agent’s prompt to determine whether the human review comment mentions issues related to its assigned dimension. If no issue is detected, the agent does not produce a refined patch. Otherwise, the agent generates a patch to address the identified issues in that dimension. After all 18 agents complete this process, the summarization LLM aggregates their outputs into a final refined patch. This adaptation preserves the original design principle of \textit{Hydra Reviewer}, where agents specialize in distinct dimensions and their outputs are summarized to produce a comprehensive result. Aside from changing the generation objective from review comments to code patches and adding human review comments as an additional input, the original prompts and system design remain unchanged.

For \textit{C-ChatGPT}, we modify its prompt to accept the patch and the human-written review comment as input, identify the issue dimensions mentioned in the comment, and generate the refined patch accordingly. The complete adapted prompts are provided in Appendix~\ref{sec:appendix-prompts}.

\subsection{ReAct Agents}
\label{sec:react-method}

The ReAct framework \cite{yao2022react} has become a widely adopted paradigm for building LLM-based agents \cite{langchain}. By combining reasoning with tool use, ReAct enables models to generate intermediate reasoning steps and interact with external tools, often improving performance and robustness compared to prompt-only approaches \cite{yao2022react,aksitov2023rest,liu2023agentbench,yang2025agentnet}. Despite its popularity, ReAct has been largely unexplored in code review automation. Given its prevalence in the LLM-agent literature, we incorporate ReAct in this study to construct both MA and SA agents.%, improving the generalizability and practical relevance of our findings.

%OLD:The ReAct method \cite{yao2022react} has gained significant popularity and adoption to build powerful LLM-based agents \cite{langchain}, which often surpass the performance and robustness of those that consist solely of prompted LLMs \cite{yao2022react, aksitov2023rest, liu2023agentbench, yang2025agentnet}. It not only elicits chains of reasoning steps from the LLMs but also allows them to interact with the environment and obtain support by invoking various tools. However, its application has been largely overlooked in code review automation tasks. Given its prevalence in the literature on LLM-based agents, we incorporated it into this study to construct agents and make our results on MA- and SA-based approaches in code review more generalizable and practical.

The tools provided for the ReAct framework fall into two categories: \textit{patch understanding} and \textit{quality-dimension assistance}. In code review, generating useful and relevant feedback requires a thorough understanding of the submitted patch \cite{ebert2019confusion}. However, the patch itself may not always provide sufficient information for this purpose \cite{li2024only,ren2025hydra,li2025consider}; additional software context is often necessary. To support patch understanding, we adopt context-retrieval tools from prior work on commit message generation \cite{li2024only,li2025consider}, a task that similarly requires a comprehensive understanding of code changes. These tools retrieve relevant contextual information associated with a patch and have been shown to improve commit message quality. We therefore employ them to assist code review automation. The full list of patch understanding tools and their functionalities described in \cite{li2024only,li2025consider} are in Appendix~\ref{sec:appendix-patchundtools}.

%The tools we provided for the ReAct method can be divided into two main categories: \textit{patch understanding} and \textit{quality dimension assistance}. In code review practice, writing a helpful, relevant review comment requires a thorough understanding of the submitted patch \cite{ebert2019confusion}. Prior research has shown that the code in the patch itself may not always contain all the necessary information to understand it \cite{li2024only,ren2025hydra, li2025consider}. Software contexts associated with the patch are highly needed. In this study, we selected the context-retrieval tools from previous works \cite{li2024only, li2025consider} used in the task of \textit{Commit Message Generation} where generating a high-quality commit message demands comprehensive code patch understanding. Since these tools have been proven to help code patch understanding to generate commit messages, we posit that they can provide valuable context information to contribute to code patch understanding that would help code review automation tasks. The complete list of these patch understanding tools and their functionalities are detailed in \cite{li2024only, li2025consider}. For more information about the selected patch understanding tools, see Appendix \ref{sec:appendix-patchundtools}.

%For example, to pinpoint the reason (described in the commit message) behind the code patch in Figure \ref{fig:example1}, the knowledge of the invoked method \textit{getComputer} is essential, which is not included in the patch. 

The software engineering community has developed numerous techniques for detecting and resolving software quality issues, which can serve as quality-dimension assistance tools for LLM agents in the ReAct framework. These tools help improve the reliability and accuracy of issue detection and resolution. Ideally, such tools would cover all 18 quality dimensions defined by \textit{Hydra Reviewer} \cite{ren2025hydra}. However, establishing a precise mapping between dimensions in the literature and Hydra Reviewer’s taxonomy is challenging and often ambiguous. For example, techniques for defect detection and automated program repair typically address both code semantic correctness and code syntax correctness \cite{li2017software,le2019automated,lee2025unidebugger}, making it difficult to assign them to a single dimension.

To cover as many dimensions as possible while considering the availability of established techniques, we reproduce six state-of-the-art methods and integrate them as quality-dimension assistance tools. These tools support functionalities including defect prediction \cite{zeng2021deep}, automated program repair \cite{xu2025aligning}, code smell detection \cite{Sharma2016}, code refactoring \cite{wu2024ismell}, code formatting style checking \cite{checkstyle}, and log statement analysis \cite{xie2024fastlog}. The complete mapping between techniques, supported dimensions, and references is provided in Appendix~\ref{sec:appendix-qualitytools}. While these tools do not cover all 18 dimensions, our goal is to compare the effectiveness of MA and SA agent architectures rather than directly compare the performance of ReAct agents against \textit{Hydra Reviewer}.

\noindent\textbf{RCG.} For the SA-based approach, we included all tools (Appendix \ref{sec:appendix-patchundtools} and \ref{sec:appendix-qualitytools}) for the agent to invoke during its reasoning and action process. We followed the best practices \cite{yao2022react,li2024only, yang2025agentnet, langchain}. The task goal in the prompts is described as generating a comprehensive review comment on the patch that identifies potential quality issues and proposes improvements. We denote this setup as \textit{SA-ReAct}. Figure \ref{fig:sa-rcg} presents the overview of \textit{SA-ReAct}. For the complete prompt, please see Appendix \ref{sec:appendix-prompts}.

Following the design principles of \textit{Hydra Reviewer}, we distributed expertise for detecting issues across multiple quality dimensions among multiple ReAct agents to construct the MA-based approach (\textit{MA-ReAct}). To do so, each ReAct agent is equipped with only one tool from the quality dimension assistance category. Since code patch understanding is a prerequisite for detecting code issues and proposing actionable solutions, all ReAct agents are equipped with tools from the patch understanding category. The complete mapping between agents and tools is presented in Appendix \ref{sec:appendix-prompts}. We acknowledge that there are tens or hundreds of distinctive MA architectures in the literature \cite{tran2025multi}, designed by researchers and practitioners to perform different tasks. Exhaustively experimenting with each one of these architectures to assemble all the ReAct agents on code review automation tasks would be infeasible. 

\begin{figure}[!htbp]
\centering
  \includegraphics[width=0.8\columnwidth]{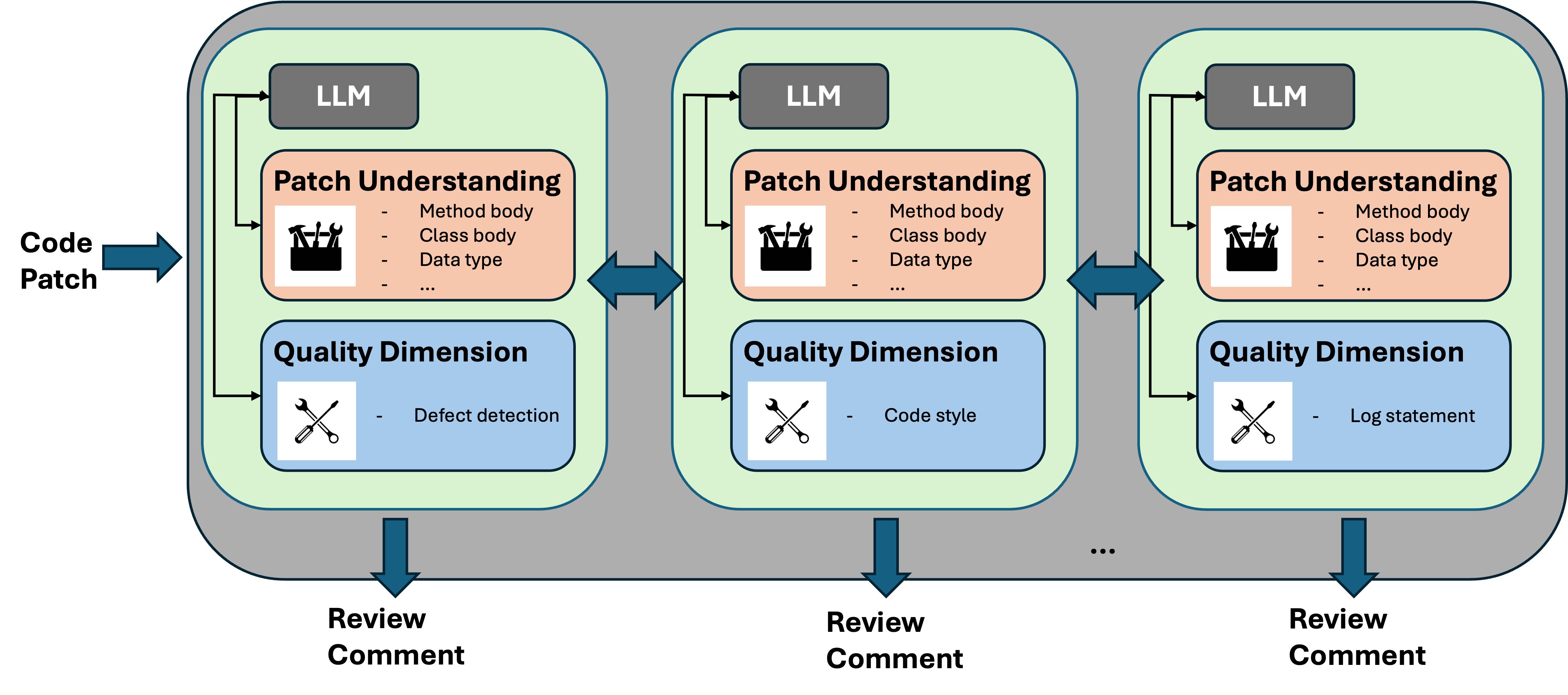}
  \caption{\textit{MA-ReAct} for RCG}
  \label{fig:ma-rcg}
\end{figure}

To build the MA architecture for agents to interact with each other, generate the final review, and ensure the best performance possible, we adopted a decentralized architecture widely adopted in practice and literature \cite{langchainhandsoff,yang2025agentnet} in which agents decide for themselves which agent to pass information and control to, and whether to generate the final code review comment. Since the quality dimensions focused by human reviewers are equally important in code review, with no one dimension more critical than another \cite{ren2025hydra}, such a design would encourage agents to treat others as peers, enabling them to receive and share information at their discretion. Additionally, different quality dimensions can be related to one another \cite{palomba2018diffuseness,hassani2018studying,abidi2021multi,panthaplackel2021deep,li2023commit}. Depending on the patches and their associated software contexts, quality issues across different dimensions can be related in various ways. Thus, by enabling agents to decide the information flow themselves, they are given greater freedom to analyze intricate relationships and act accordingly. Moreover, this design avoids the drawbacks commonly observed in MA systems with centralized coordination by a supervisor agent, such as scalability bottlenecks, limited adaptability, and single points of failure \cite{yang2025agentnet, dang2025multi}. The overview of \textit{MA-ReAct} for RCG is shown in Figure \ref{fig:ma-rcg}.

\begin{figure}[!htbp]
`\centering
  \includegraphics[width=0.65\columnwidth]{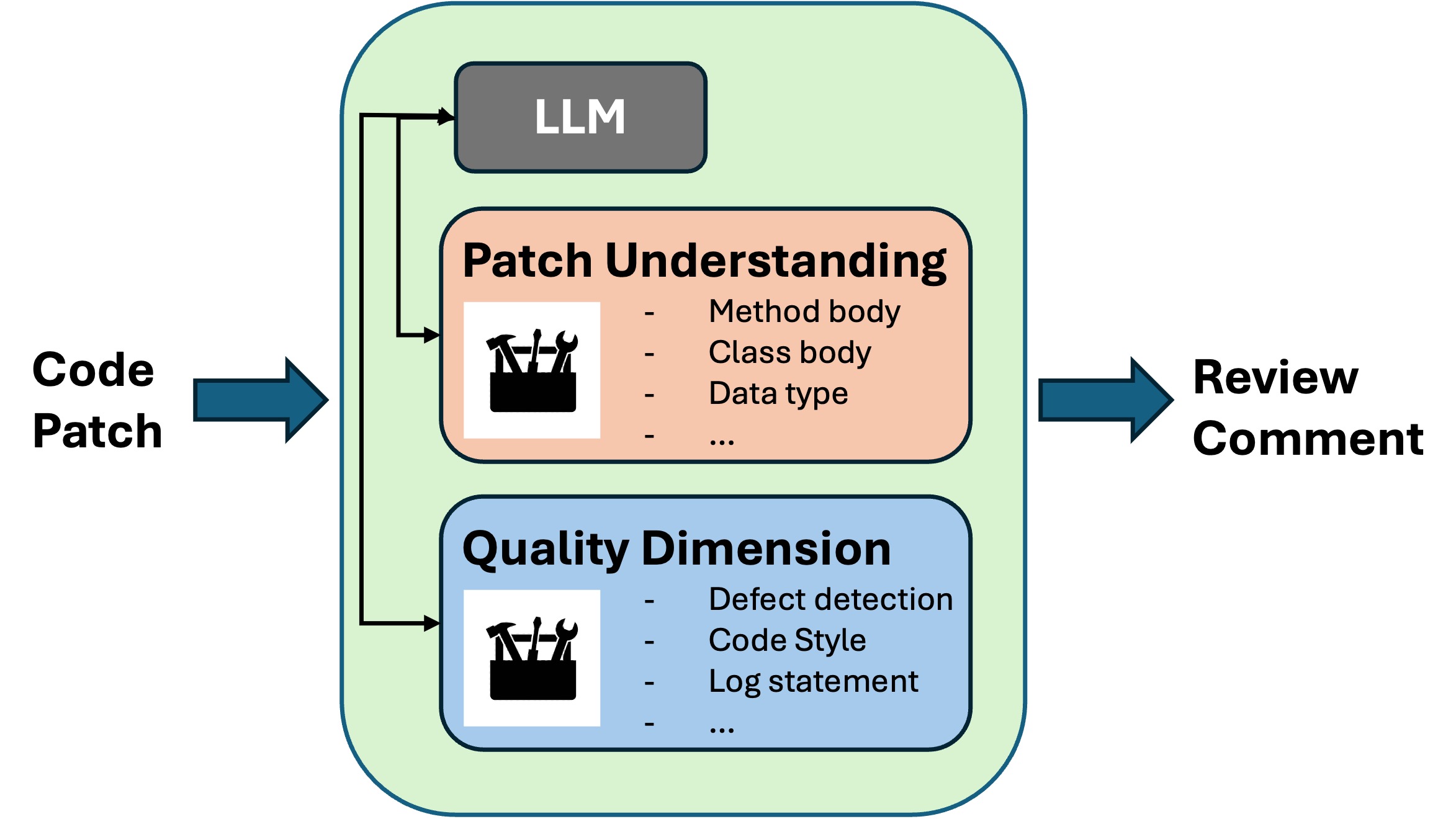}
  \caption{\textit{SA-ReAct} for RCG}
  \label{fig:sa-rcg}
\end{figure}

\begin{figure}[!htbp]
\centering
  \includegraphics[width=0.65\columnwidth]{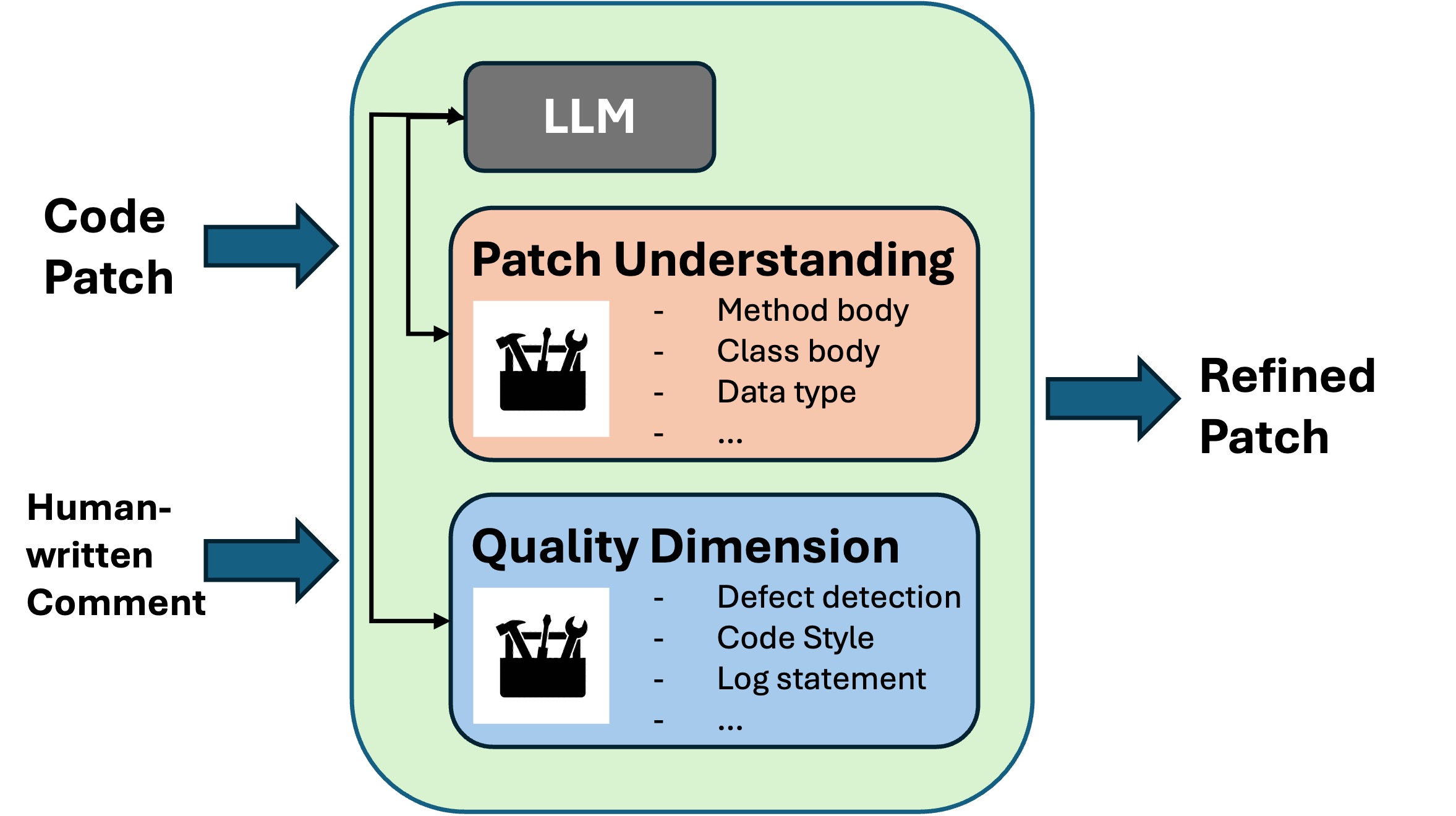}
  \caption{\textit{SA-ReAct} for CR}
  \label{fig:sa-cr}
\end{figure}

\begin{figure}[!htbp]
\centering
  \includegraphics[width=0.8\columnwidth]{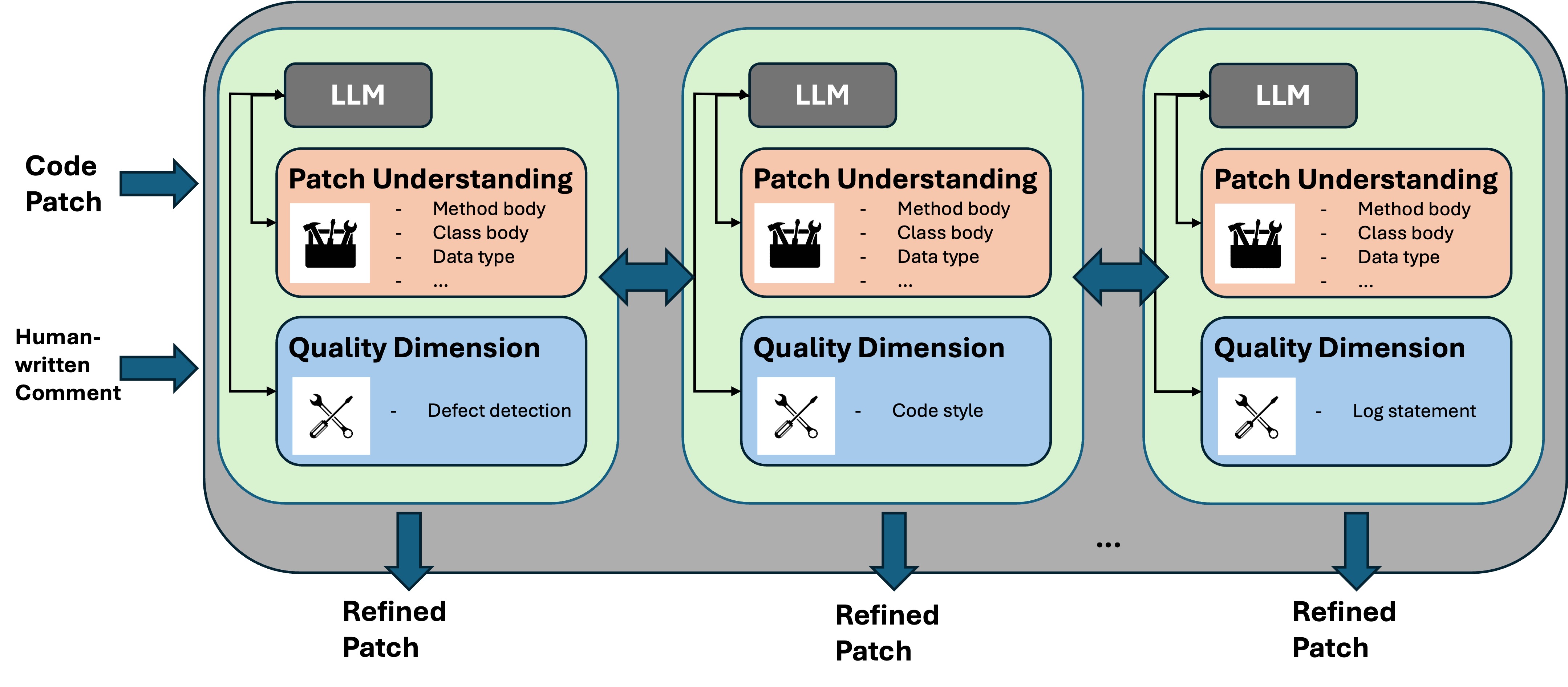}
  \caption{\textit{MA-ReAct} for CR}
  \label{fig:ma-cr}
\end{figure}

\noindent\textbf{CR.} Similar to that of \textit{Hydra Reviewer}, we fed \textit{ChatGPT-CR} with the review comments generated by \textit{SA-ReAct} and \textit{MA-ReAct} on RCG to obtain the refined code patches. Also, we investigated the effectiveness of these two agentic systems in refining the patches by themselves. To do so, given human-written review comments and submitted code patches as input, we revised the prompts for \textit{SA-ReAct} and \textit{MA-ReAct} to set the goal of generating a refined patch rather than review comments. Figure \ref{fig:sa-cr} and \ref{fig:ma-cr} show the overview of these two approaches. Appendix \ref{sec:appendix-prompts} presents the detailed prompts used in our experiments. 

% To produce a PDF file, pdf\LaTeX{} is strongly recommended (over original \LaTeX{} plus dvips+ps2pdf or dvipdf).
% The style file \texttt{acl.sty} can also be used with
% lua\LaTeX{} and
% Xe\LaTeX{}, which are especially suitable for text in non-Latin scripts.
% The file \texttt{acl\_lualatex.tex} in this repository provides
% an example of how to use \texttt{acl.sty} with either
% lua\LaTeX{} or
% Xe\LaTeX{}.

\section{Experimental Setup}

\subsection{Model Selection} 
Our goal is to evaluate the effectiveness of MA- and SA-based approaches on RCG and CR. Since both rely on an LLM backbone, we use the same model across all systems to ensure differences arise from the agent architecture rather than the underlying model. Following \textit{Hydra Reviewer}, we use \textit{GPT-4o-mini} \cite{gpt4} as the backbone. We set the temperature to 0 to obtain the most deterministic outputs \cite{liu2023your,li2024only}. 

%OLD: The goal of this study is to investigate the effectiveness of MA- and SA-based approaches in code review automation tasks, namely, RCG and CR. Since the reasoning and generation backbones of these approaches are an LLM, we set the LLM used by all approaches in our experiments to the same one. Thus, any notable performance difference would stem from the agent architecture rather than the underlying LLM. To select the backbone model for all the agent architectures, we adopted \textit{GPT-4o-mini} \cite{gpt4} that is also used by \textit{Hydra Reviewer}. 

\subsection{Cost Analysis}
To investigate the computational overhead incurred by MA- and SA-based approaches, we calculated the number of tokens processed by GPT-4o-mini using \textit{tiktoken} \cite{tiktoken}. Specifically, we considered all tokens that were input into the LLM and all output tokens from the LLM when performing RCG or CR task for a single data instance. For ReAct-based methods, we also collected the number of tool calls throughout the task completion.

\subsection{Evaluation}

In this section, we present the evaluation methods to measure the performance of MA- and SA-based approaches in RCG and CR, respectively. 

\subsubsection{RCG Task Evaluation}
\label{sec:RCGevalMetric}
Most prior studies on automated RCG evaluate performance using text similarity metrics between generated comments and human-authored comments (i.e., ground truth), such as BLEU \cite{papineni2002bleu} \cite{li2022automating,hong2022commentfinder,guo2024exploring,ren2025hydra}. However, recent work \cite{liu2025too,sghaier2025harnessing,ren2025hydra} shows that human-authored review comments often contain quality issues, including unclear or irrelevant content, noisy feedback without actionable suggestions, redundant information, and a lack of detection of certain issues. Comparing generated comments against such imperfect references may unfairly penalize high-quality outputs that provide clearer explanations or more relevant suggestions. Therefore, we adopt a set of reference-free evaluation metrics that directly assess the quality of generated review comments, enabling a more comprehensive and fair evaluation. It's worth noting that we directly used the LLM-as-a-judge implementations from the metrics' replication packages, which have been validated and assessed by the researchers in their studies.

%OLD: To evaluate the performance of RCG automation techniques, most existing code review automation studies \cite{li2022automating,hong2022commentfinder,guo2024exploring, ren2025hydra} have adopted metrics that measure text similarity between the generated comments and human-authored comments (i.e., ground truth), such as BLEU score \cite{papineni2002bleu}. 

%OLD: However, recent research \cite{liu2025too,sghaier2025harnessing,ren2025hydra} has highlighted that human-authored review comments contain quality issues. These issues include comments that are unclear, unreadable, or irrelevant to the code patches, noisy comments that do not propose any actionable suggestions, and comments that are full of redundant information or lack the detection of certain issues. Directly comparing generated comments with such low-quality ground truth may unfairly affect high-quality generated comments with clear suggestions and relevant explanations. Therefore, we adopted a range of quality evaluation metrics that comprehensively measure the quality of the generated review comments without any human-written references to ensure a comprehensive and fair evaluation. 

\noindent \textbf{Conciseness} measures how briefly a review comment conveys necessary information without redundant explanations. Following \cite{sghaier2025harnessing}, we use an LLM-as-a-judge to score each comment on a 1–10 scale, where 1 indicates the most verbose and 10 the most concise.

%OLD:\textbf{Conciseness} measures the brevity of the review comments and the degree to which the comment includes only necessary information without redundant explanation. The LLM-as-a-judge, as implemented by \cite{sghaier2025harnessing}, is used to evaluate review comments for this metric. It outputs scores from 1 to 10, with 1 being the most verbose and 10 the most concise. 

\noindent \textbf{Clarity} measures how clearly and readably a review comment conveys its message. Following \cite{sghaier2025harnessing}, we use an LLM-as-a-judge to score each comment on a 1–10 scale, where 1 indicates the least clear and 10 the most clear.

%\textbf{Clarity} measures how clear and readable the review comment conveys its message. The LLM-as-a-judge implemented by \cite{sghaier2025harnessing} is utilized to evaluate review comments in this metric. It outputs scores from 1 to 10, with 1 being the least clear and 10 the most clear.

\noindent \textbf{Relevance} measures how closely a review comment relates to the code patch. Following \cite{sghaier2025harnessing}, we use an LLM-as-a-judge to score each comment on a 1–10 scale, where 1 indicates completely irrelevant or off-topic content, and 10 indicates high relevance.
%\textbf{Relevance} assesses the extent to which the review comment is relevant to the code patch. The LLM-as-a-judge implemented by \cite{sghaier2025harnessing} is utilized to evaluate review comments in this metric. It outputs scores from 1 to 10, with 1 being completely irrelevant and off-topic and 10 being highly relevant.

\noindent\textbf{Accuracy} evaluates whether a review comment correctly identifies issues without introducing errors or irrelevant information. Following the evaluation protocol of \cite{jaoua2025combining}, we adopt an LLM-as-a-judge that classifies each review comment into one of three categories: \textit{accurate}, \textit{partially accurate}, or \textit{not accurate}. A review is labeled \textit{accurate} if it correctly identifies code quality issues without errors or irrelevant content. It is labeled \textit{partially accurate} if some parts are valid while others are incorrect or irrelevant. A review is considered \textit{not accurate} if it fails to identify the issues and is completely irrelevant to the context.

%OLD: evaluates whether the review comment correctly addresses issues without any errors or irrelevant information. Following the evaluation practice by \cite{jaoua2025combining} and their LLM-as-a-judge, the evaluation of this metric is to classify the review comment into one of the three categories: \textit{accurate}, \textit{partially accurate}, or \textit{not accurate}. A review is considered accurate if it correctly identifies code quality issues without any errors or irrelevant information. If part of the review is valid, and the rest is irrelevant or incorrect, it is considered partially accurate. Finally, a review is classified as not accurate if it fails to mention the identified issues and is completely irrelevant to the context.

\noindent\textbf{Validity} assesses whether a review comment is \textit{valid} or \textit{noisy}. Following the evaluation protocol of \cite{liu2025too}, we use an LLM-as-a-judge to classify each comment into these two categories. A comment is considered \textit{valid} if it explicitly requests code changes intended to improve the patch. In contrast, a \textit{noisy} comment does not suggest actionable changes or improvements to the code.

%OLD: determines whether the review comment is valid or noisy. Following the evaluation practice by \cite{liu2025too}, and their LLM-as-a-judge, a valid comment refers to a comment requesting explicit code changes aimed at improving the code patch, and a noisy comment does not request code changes or actions to improve the code quality. 

% \noindent\textbf{Text Similarity} includes metrics that compare the textual similarity between the generated review comment and the human-written comment as the ground truth. Since they are widely used in studies of code review automation \cite{li2022automating,hong2022commentfinder,guo2024exploring, ren2025hydra}, we also adopted them in this work. Specifically, we include \textit{BLEU} \cite{papineni2002bleu}, \textit{Meteor} \cite{banerjee2005meteor}, and \textit{Rouge-L} \cite{lin2004rouge}. 

% \noindent\textbf{Coverage} measures the number of code quality dimensions covered by the review comment. 

\subsubsection{CR Task Evaluation}
\label{sec:CR_eval_metrics}

To evaluate CR, we compare the generated refined patches with human-refined patches using similarity-based metrics. We also employ an LLM-as-a-judge to assess the quality of the refined patches across the code quality dimensions commonly considered in code review.

%To evaluate the quality of the refined code in the task of CR, we compared the generated refined code patches with human-refined patches. To complement the similarity-based evaluation, we also built an LLM-as-a-judge to directly evaluate the quality of the refined patches across quality dimensions that practitioners focus on when conducting code review. 

\noindent \textbf{Code Similarity} measures the similarity between the generated refined patch and the human-authored refined patch. We use \textit{CodeBLEU} \cite{ren2020codebleu}, a metric widely used \cite{pornprasit2024fine,ben2024improving,sghaier2025harnessing}.
%\textbf{Code Similarity} measures the similarity between the generated refined patch and the human-authored refined patch. In this study, we selected \textit{CodeBLEU} \cite{ren2020codebleu} because it has been widely adopted in previous CR automation work \cite{pornprasit2024fine,ben2024improving,sghaier2025harnessing}. 

\noindent\textbf{Code Quality} assesses the quality of the refined patch by estimating the number of code quality issues it contains. We evaluate issues across the 18 quality dimensions identified by \textit{Hydra Reviewer}, which reflect dimensions commonly considered during code review. To implement this evaluation, we adapted the prompts from the 18 agents in \textit{Hydra Reviewer} to build 18 LLM-as-a-judges, each responsible for identifying the number of issues within a specific dimension and providing a rationale. The judges are constructed following best practices for LLM-based evaluation \cite{pan2024human,sghaier2025harnessing,jaoua2025combining,liu2025too,crupi2025effectiveness}. The full prompts are provided in Appendix~\ref{sec:appendix-prompts}.

%\hl{Should we evaluate these judges like what I wrote here or just say we built them following the 18 agents and best practices? Would it be too much if we evaluate them by manually checking for the issues?} 

% \hl{I think this is good, not too much}

To validate the reliability of these judges, two researchers with more than eight years of Java development experience manually annotated the number of issues across all 18 dimensions for 100 randomly sampled patches. Following prior work \cite{liu2023g,wang2023chatgpt,crupi2025effectiveness}, we computed Pearson correlation coefficients \cite{sedgwick2012pearson} between the judges’ outputs and the human annotations (the average of issue numbers from the two researchers). The correlations range from 0.395 to 0.684, indicating moderate to strong agreement. Detailed results for all dimensions are reported in Appendix~\ref{sec:appendix-judge}. % \hl{the appendix is not complete}

\subsection{Dataset}
We use the \textit{CuRev} dataset \cite{sghaier2025harnessing}, which has been widely adopted in prior studies on automated code review \cite{li2022automating,lu2023llama,ben2024improving}. The dataset spans nine programming languages, and each instance contains a submitted code patch, a human-written review comment, and a human-authored refined patch. In this study, we focus on Java because the patch-understanding tools used in our approach (Section~\ref{sec:react-method}) support only Java; extending them to other languages is beyond the scope of this work. Due to the computational and financial cost of running the automation systems and LLM-based evaluators, we randomly sample 328 instances (95\% confidence level, 5\% margin of error) for evaluation.

\section{Experimental Result Analysis}
\label{sec:results}

This section discusses the performance of MA- and SA-based approaches in the two code review automation tasks.

\subsection{RCG Analysis}

% \begin{table*}[ht]
% \centering
% \caption{The Evaluation Results of All Techniques for RCG. We report the average scores across the instances in our dataset for Conciseness, Clarity, and Relevance. For Accuracy, we report the percentages of comments that are accurate/partially accurate. For Validity, we report the percentage of valid comments.}
% \resizebox{\linewidth}{!}{%
% \begin{tabular}{l|cccccccc}
% \hline
% \textbf{}               & \textbf{Conciseness} & \textbf{Clarity} & \textbf{Relevance} & \textbf{Accuracy} & \textbf{Validity} & \textbf{BLEU} & \textbf{METEOR} & \textbf{ROUGE-L} \\ \hline
% \textbf{Hydra Reviewer} & 6.28                 & 7.97             & 8.27               & 48.6\%/21.0\%     & 89.3\%            & 0.83          & 4.9             & 3.83             \\
% \textbf{C-ChatGPT}      & 7.42                 & 8.68             & 8.72               & 55.8\%/15.4\%     & 94.7\%            & 1.55          & 6.83            & 5.87             \\ \hline
% \textbf{MA-ReAct}       & 6.68                 & 8.22             & 8.23               & 74.9\%/11.0\%     & 79.6\%            & 1.30          & 6.45            & 4.53             \\
% \textbf{SA-ReAct}       & 6.87                 & 8.61             & 8.56               & 63.6\%/25.7\%     & 97.8\%            & 0.87          & 5.42            & 3.85             \\ \hline
% \end{tabular}
% }
% \label{tab:eval_RCG}
% \end{table*}

\begin{table}[ht]
\centering
\caption{The evaluation results of all techniques for RCG. We report the average scores across the instances in our dataset for Conciseness, Clarity, and Relevance. For Accuracy, we report the percentages of comments that are accurate/partially accurate. For Validity, we report the percentage of valid comments.}
\resizebox{\linewidth}{!}{%
\begin{tabular}{l|ccccc}
\hline
\textbf{}               & \textbf{Conciseness} & \textbf{Clarity} & \textbf{Relevance} & \textbf{Accuracy} & \textbf{Validity} \\ \hline
\textbf{Hydra Reviewer} & 6.28                 & 7.97             & 8.27               & 48.6\%/21.0\%     & 89.3\%            \\
\textbf{C-ChatGPT}      & 7.42                 & 8.68             & 8.72               & 55.8\%/15.4\%     & 94.7\%            \\ \hline
\textbf{MA-ReAct}       & 6.68                 & 8.22             & 8.23               & 74.9\%/11.0\%     & 79.6\%            \\
\textbf{SA-ReAct}       & 6.87                 & 8.61             & 8.56               & 63.6\%/25.7\%     & 97.8\%            \\ \hline
\end{tabular}
}
\label{tab:eval_RCG}
\end{table}

Table~\ref{tab:eval_RCG} reports the evaluation results using the metrics described in Section~\ref{sec:RCGevalMetric}. For \textit{Hydra Reviewer} and its SA counterpart \textit{C-ChatGPT}, the results show that \textit{C-ChatGPT} achieves statistically significantly higher scores in Conciseness, Clarity, and Relevance. We included the results of the statistical analyses in Appendix \ref{sec:appendix-evalresults}. \textit{C-ChatGPT} also produces a higher proportion of Accurate (55.8\%) and Valid (94.7\%) comments than \textit{Hydra Reviewer}. These results indicate that the SA-based \textit{C-ChatGPT} can generate higher-quality review comments despite the more complex multi-agent design of \textit{Hydra Reviewer}.

A similar trend is observed for the ReAct-based approaches: \textit{SA-ReAct} outperforms \textit{MA-ReAct} in Conciseness, Clarity, and Relevance with statistical significance (Appendix \ref{sec:appendix-evalresults}). Although \textit{SA-ReAct} produces slightly less Accurate comments (63.6\%), the majority of its outputs are Valid (97.8\%), substantially higher than \textit{MA-ReAct} (79.6\%).

%OLD: Table \ref{tab:eval_RCG} shows the evaluation results measured by the metrics introduced in Section \ref{sec:RCGevalMetric}. For \textit{Hydra Reviewer} and its SA-based counterpart \textit{C-ChatGPT}, the results reveal that \textit{C-ChatGPT} generates better review comments than \textit{Hydra Reviewer} in Conciseness, Clarity, and Relevance. In addition, more comments generated by \textit{C-ChatGPT} are accurate (55.8\%) and valid (94.7\%) than those by \textit{Hydra Reviewer}. Evidently, Table \ref{tab:eval_RCG} highlights that, while \textit{C-ChatGPT} is a SA-based approach, it can still generate better code review comments than \textit{Hydra Reviewer}, which consists of multiple agents and various assistant units. Similarly, \textit{SA-ReAct} generates better review comments than \textit{MA-ReAct} in Conciseness, Clarity, and Relevance. While \textit{SA-ReAct} does fall slightly behind in the percentage of accurate comments (63.6\%), the majority of its generated comments (97.8\%) turn out to be valid, which is significantly higher than that of the \textit{MA-ReAct} (79.6\%). 

Overall, the results show that MA approaches, despite their greater architectural complexity and specialized agents, do not produce review comments that are more Concise, Clear, Relevant, or Accurate than those generated by SA-based approaches. In fact, SA approaches outperform MA-based ones on most metrics. These findings challenge the common assumption that MA approaches inherently benefit RCG by distributing responsibility across specialized agents. Instead, a well-designed SA approach covering all quality dimensions can match or even surpass MA approaches in generating high-quality review comments.

%OLD: Overall, the results show that while MA-based approaches are more complex and involve multiple agents with different expertise collaborating, they do not generate review comments that are more concise, clear, relevant, or accurate than those generated by SA-based approaches. On the contrary, SA-based approaches outperform MA-based ones in almost all metrics. These findings contradict the belief in the literature that MA can benefit RCG where detecting issues of different dimensions is distributed among multiple agents. A single agent that covers all dimensions can detect the issues and generate review comments as good as an MA-based system, if not better. 

\subsection{CR Analysis}

\begin{table*}[ht]
\caption{Code quality evaluation results of \textit{Hydra Reviewer} and \textit{C-ChatGPT} for CR. The refined patches are generated by ChatGPT-CR based on comments from different techniques. We report the average number of quality issues detected by the LLM-as-a-judge. Grey cells indicate statistically significant differences.}
\centering
\resizebox{\linewidth}{!}{%
\begin{tabular}{l|cccccc}
\hline
                        & \textbf{Formatting Style}           & \textbf{Logic Readability}   & \textbf{Semantic Correctness}             & \textbf{Syntax Correctness}  & \textbf{Testing}                       & \textbf{Comment Quality}         \\ \hline
\textbf{Hydra Reviewer} & \cellcolor[HTML]{C0C0C0}2.54        & \cellcolor[HTML]{C0C0C0}1.75 & \cellcolor[HTML]{C0C0C0}1.64              & 0.83 & \cellcolor[HTML]{C0C0C0}2.46           & \cellcolor[HTML]{C0C0C0}3.69     \\
\textbf{C-ChatGPT}      & \cellcolor[HTML]{C0C0C0}2.22        & \cellcolor[HTML]{C0C0C0}1.32 & \cellcolor[HTML]{C0C0C0}0.94              & 0.74 & \cellcolor[HTML]{C0C0C0}1.8            & \cellcolor[HTML]{C0C0C0}2.69     \\ \hline
                        & \textbf{Comment Style}              & \textbf{Compatibility}       & \textbf{Extensibility}                    & \textbf{Fault Tolerance}     & \textbf{Identifier Naming Readability} & \textbf{Identifier Naming Style} \\ \hline
\textbf{Hydra Reviewer} & \cellcolor[HTML]{C0C0C0}3.61        & \cellcolor[HTML]{C0C0C0}1.36 & \cellcolor[HTML]{C0C0C0}2.50              & 2.15                         & \cellcolor[HTML]{C0C0C0}1.99           & \cellcolor[HTML]{C0C0C0}2.42     \\
\textbf{C-ChatGPT}      & \cellcolor[HTML]{C0C0C0}2.62        & \cellcolor[HTML]{C0C0C0}0.96 & \cellcolor[HTML]{C0C0C0}2.00              & 2.22                         & \cellcolor[HTML]{C0C0C0}1.52           & \cellcolor[HTML]{C0C0C0}2.07     \\ \hline
                        & \textbf{Name and Logic Consistency} & \textbf{Performance}         & \textbf{Programming Handling Conventions} & \textbf{Redundancy}          & \textbf{Runtime Observability}         & \textbf{Security Compliance}     \\ \hline
\textbf{Hydra Reviewer} & \cellcolor[HTML]{C0C0C0}1.39        & \cellcolor[HTML]{C0C0C0}1.68 & \cellcolor[HTML]{C0C0C0}2.53              & \cellcolor[HTML]{C0C0C0}1.86 & \cellcolor[HTML]{C0C0C0}2.47           & 0.5                              \\
\textbf{C-ChatGPT}      & \cellcolor[HTML]{C0C0C0}1.06        & \cellcolor[HTML]{C0C0C0}1.29 & \cellcolor[HTML]{C0C0C0}1.95              & \cellcolor[HTML]{C0C0C0}1.40 & \cellcolor[HTML]{C0C0C0}1.80           & 0.46                             \\ \hline
\end{tabular}
}
\label{tab:llmeval_hydra_generatedcomment}
\end{table*}

\begin{table*}[ht]
\caption{Code Quality Evaluation Results of adapted \textit{Hydra Reviewer} and \textit{C-ChatGPT} for CR. Refined patches are generated by these adapted techniques based on comments from human reviewers. We report the average number of quality issues detected by the LLM-as-a-judge. Grey cells indicate statistically significant differences.}
\centering
\resizebox{\linewidth}{!}{%
\begin{tabular}{l|cccccc}
\hline
                        & \textbf{Formatting Style}           & \textbf{Logic Readability}   & \textbf{Semantic Correctness}             & \textbf{Syntax Correctness}  & \textbf{Testing}                       & \textbf{Comment Quality}         \\ \hline
\textbf{Hydra Reviewer} & \cellcolor[HTML]{C0C0C0}2.34        & \cellcolor[HTML]{C0C0C0}1.45 & \cellcolor[HTML]{FFFFFF}0.76              & \cellcolor[HTML]{C0C0C0}0.62 & \cellcolor[HTML]{C0C0C0}1.84           & \cellcolor[HTML]{C0C0C0}3.04     \\
\textbf{C-ChatGPT}      & \cellcolor[HTML]{C0C0C0}1.91        & \cellcolor[HTML]{C0C0C0}1.15 & \cellcolor[HTML]{FFFFFF}0.82              & \cellcolor[HTML]{C0C0C0}0.77 & \cellcolor[HTML]{C0C0C0}1.47           & \cellcolor[HTML]{C0C0C0}2.57     \\ \hline
                        & \textbf{Comment Style}              & \textbf{Compatibility}       & \textbf{Extensibility}                    & \textbf{Fault Tolerance}     & \textbf{Identifier Naming Readability} & \textbf{Identifier Naming Style} \\ \hline
\textbf{Hydra Reviewer} & \cellcolor[HTML]{C0C0C0}3.04        & \cellcolor[HTML]{FFFFFF}0.80 & \cellcolor[HTML]{C0C0C0}1.95              & \cellcolor[HTML]{C0C0C0}2.54 & \cellcolor[HTML]{C0C0C0}1.77           & \cellcolor[HTML]{C0C0C0}2.20     \\
\textbf{C-ChatGPT}      & \cellcolor[HTML]{C0C0C0}2.35        & \cellcolor[HTML]{FFFFFF}0.88 & \cellcolor[HTML]{C0C0C0}1.67              & \cellcolor[HTML]{C0C0C0}2.31 & \cellcolor[HTML]{C0C0C0}1.34           & \cellcolor[HTML]{C0C0C0}1.79     \\ \hline
                        & \textbf{Name and Logic Consistency} & \textbf{Performance}         & \textbf{Programming Handling Conventions} & \textbf{Redundancy}          & \textbf{Runtime Observability}         & \textbf{Security Compliance}     \\ \hline
\textbf{Hydra Reviewer} & \cellcolor[HTML]{FFFFFF}0.93        & \cellcolor[HTML]{C0C0C0}1.35 & \cellcolor[HTML]{C0C0C0}1.62              & \cellcolor[HTML]{C0C0C0}1.54 & \cellcolor[HTML]{C0C0C0}1.96           & 0.41                             \\
\textbf{C-ChatGPT}      & \cellcolor[HTML]{FFFFFF}0.91        & \cellcolor[HTML]{C0C0C0}1.08 & \cellcolor[HTML]{C0C0C0}1.97              & \cellcolor[HTML]{C0C0C0}1.19 & \cellcolor[HTML]{C0C0C0}1.55           & 0.36                             \\ \hline
\end{tabular}
}
\label{tab:llmeval_hydra_humancomment}
\end{table*}

\begin{table}[]
\caption{Code similarity evaluation results for CR. The refined patches are generated by ChatGPT-CR based on comments from different techniques.}
\resizebox{\columnwidth}{!}{%
\begin{tabular}{lcc|cc}
\hline
                  & \textbf{Hydra Reviewer} & \textbf{C-ChatGPT} & \textbf{MA-ReAct} & \textbf{SA-ReAct} \\ \hline
\textbf{CodeBLEU} & 0.41                    & 0.40               & 0.42              & 0.42              \\ \hline
\end{tabular}
}
\label{tab:eval_codebleu_generatedcomments}
\end{table}

\begin{table}[]
\caption{Code similarity evaluation results for CR. The refined patches are generated by adapted agent-based techniques based on comments from human reviewers.}
\resizebox{\columnwidth}{!}{%
\begin{tabular}{lcc|cc}
\hline
                  & \textbf{Hydra Reviewer} & \textbf{C-ChatGPT} & \textbf{MA-ReAct} & \textbf{SA-ReAct} \\ \hline
\textbf{CodeBLEU} & 0.43                    & 0.44               & 0.42              & 0.42              \\ \hline
\end{tabular}
}
\label{tab:eval_codebleu_humancomments}
\end{table}

We evaluate CR under two experimental setups. In the first setup, the state-of-the-art technique \textit{ChatGPT-CR} is provided with code patches and the review comments generated by the MA- and SA-based approaches on RCG in our study to produce refined patches. In the second setup, we adapt the MA- and SA-based approaches on RCG to directly refine patches using human-written review comments. The evaluation results are presented in this section; additional details on the experimental design are provided in Section~\ref{sec:methodology}.

%To conduct the CR task, we have two experimental setups. First, the state-of-the-art CR technique, \textit{ChatGPT-CR}, is input with code patches along with the review comments generated by MA- and SA-based techniques in this study to produce refined patches. In the second setup, we adapted MA- and SA-based techniques to refine patches based on human reviewers' original code review comments. We present the evaluation results in this section. More details about the experimental design can be found in Section \ref{sec:methodology}.

% \hl{table numbers are messed up, RCG analysis starts with table 4, should be table 1, then as we progress table numbers should increase, right now it's reversed, fix it}

\noindent\textbf{CR on Generated Comments.} Table~\ref{tab:eval_codebleu_generatedcomments}, showing that refined patches generated by \textit{ChatGPT-CR} using review comments from different techniques, exhibits similar CodeBLEU scores when compared to human-refined patches. However, Table~\ref{tab:eval_RCG} indicates that the quality of review comments varies across techniques. To better assess the impact of these comments, we further evaluate the refined patches using the LLM-as-a-judge across 18 quality dimensions (Section~\ref{sec:CR_eval_metrics}).

Comparing \textit{Hydra Reviewer} with its SA counterpart \textit{C-ChatGPT} (Table~\ref{tab:llmeval_hydra_generatedcomment}), patches refined based on \textit{Hydra Reviewer}-generated comments contain significantly more quality issues across 15 dimensions than those refined using \textit{C-ChatGPT}. This suggests that a simpler SA approach can produce review comments that better support high-quality refinement than the more complex MA-based design. For the ReAct-based methods, we found that \textit{MA-ReAct} and \textit{SA-ReAct} have comparable effects on \textit{ChatGPT-CR}. Due to space constraints, we include the results of the ReAct-based methods and all statistical tests analyzing the score differences in Appendix \ref{sec:appendix-evalresults}. Overall, these results indicate that distributing expertise across multiple agents does not yield clear advantages over a single-agent architecture in this setting.

% \hl{for the react based method, we can simply sya shows similar pattern and is in appendix dur to space contsraints}
% For the ReAct-based methods, Table~\ref{tab:llmeval_react_generatedcomment} shows that \textit{MA-ReAct} and \textit{SA-ReAct} have comparable effects on \textit{ChatGPT-CR}, as differences across most dimensions are not statistically significant (Welch’s t-test \cite{ruxton2006unequal}). Overall, these results indicate that distributing expertise across multiple agents does not yield clear advantages over a single-agent architecture in this setting.

%\hl{I need to discuss something somewhere, such as why we expect MA would have better performance, what is the extra cost by MA compared to SA etc. }

\noindent\textbf{CR on Human-written Comments.} 
To evaluate the ability of MA- and SA-based approaches to directly refine code patches using human-written review comments, we adapt all selected techniques to generate refined patches instead of review comments (Section~\ref{sec:methodology}). As shown in Table~\ref{tab:eval_codebleu_humancomments}, CodeBLEU scores are similar across all techniques, indicating that similarity-based metrics alone do not capture differences in patch quality. We therefore further evaluate the refined patches using the LLM-as-a-judge across 18 quality dimensions (Tables~\ref{tab:llmeval_hydra_humancomment})

%and \ref{tab:llmeval_react_humancomment} \hl{I think we should put the react table in appendix}).

Compared with \textit{Hydra Reviewer}, patches refined by \textit{C-ChatGPT} contain significantly fewer issues in 14 of the 18 dimensions. Similarly, \textit{SA-ReAct} achieves comparable performance to \textit{MA-ReAct} in 16 dimensions and produces fewer issues in the remaining two (Appendix \ref{sec:appendix-evalresults}). Despite involving only a single agent, these results suggest that SA-based approaches perform as well as, or better than, MA-based approaches for the CR task.

\subsection{Cost Comparison}

We calculated the approaches' computational overhead in token consumption on RCG and CR. For CR, we only calculated token consumption in the setting where adapted MA and SA-based approaches themselves generated refined patches based on human-written review comments, given that the token cost by \textit{ChatGPT-CR} would only be the input comments and output refined patches. On average, \textit{C-ChatGPT} used 1,610 tokens on RCG and 1,645 on CR for each data instance, while \textit{Hydra Reviewer} had 47,222 on RCG and 47,627 on CR. For ReAct-based methods, \textit{SA-ReAct} processed 3,667 tokens on RCG and 2,571 on CR, while \textit{MA-ReAct} used 35,031 on RCG and 50,407 on CR. Also, \textit{MA-ReAct} involved significantly more tool calls. Due to space constraints, we include the full cost analysis statistics in Appendix \ref{sec:appendix-cost}. The results reveal that SA-based approaches are more cost-efficient than MA-based approaches in code review automation tasks.

\section{Related Work}
\label{sec:bibtex}

%Our work provides valuable insights into the research of code review automation involving LLM-based agents. Regarding the related works, 

%New
\noindent\textbf{Code Review Automation.} Guo et al. \cite{guo2024exploring} studied the use of ChatGPT to refine code based on review comments. Lu et al. \cite{lu2023llama} explored automated code review by fine-tuning LLaMA with parameter-efficient methods. Because code review involves identifying issues across multiple quality dimensions and often requires collaboration among developers with different roles, recent work has increasingly explored MA approaches. For example, Ren et al. \cite{ren2025hydra} proposed \textit{Hydra Reviewer}, which employs 18 agents specializing in different code-quality dimensions to analyze patches and generate review comments. Tang et al. \cite{tang2024codeagent} introduced \textit{CodeAgent}, where agents assume distinct roles to generate review comments and refine code.

\noindent\textbf{Single-Agent VS Multi-Agent.} While MA approaches have achieved strong results across many tasks \cite{tang2024codeagent,ren2025hydra,lee2025unidebugger,wang2025tdag}, recent work reports mixed evidence regarding their advantages over SA approaches. Xu et al. \cite{xu2026rethinking} show that a strong SA approach that uses multi-turn interaction and a shared KV cache can match or outperform a homogeneous MA approach while being more efficient. Wang et al. \cite{wang2024rethinking} show that appropriately prompted SA models perform comparably to MA discussion frameworks on reasoning tasks, with MA benefits appearing mainly when SA models are weakly prompted. Similarly, Cemri et al. \cite{cemri2025multi} find that MA approaches often fail to significantly outperform strong SA baselines and may suffer from coordination failures and quality control issues. Gao et al. \cite{gao2025single} show that MA-SA performance gap narrows on complex agentic tasks with modern LLMs. However, whether these observations hold for code review automation tasks remains unclear. This work addresses that gap. %Although code review automation tasks appear well-suited for an MA approach, in which agents specialize in different quality dimensions or roles, we show that SA approaches can perform as effectively as state-of-the-art MA systems.

\section{Conclusion}

In this paper, we evaluate the performance of state-of-the-art MA- and SA-based techniques on RCG and CR. Although these tasks are widely assumed to benefit from MA-based approaches, our results show that SA-based approaches perform comparably to—and in some cases outperform—MA approaches. These findings underscore the need to carefully evaluate the trade-offs between MA and SA designs, rather than defaulting to MA architectures for complex software engineering tasks. %Future work should examine these trade-offs across a broader range of software engineering tasks.

%Our findings highlight the importance of carefully assessing the trade-offs between MA and SA designs, rather than defaulting to MA approaches for complex software engineering tasks. Future research should analyze the trade-off between MA- and SA-based approaches on other software engineering tasks. 

%OLD: While these tasks are believed to benefit most from MA-based approaches, our results demonstrate that, surprisingly, SA-based approaches are as effective as MA-based ones, and sometimes even better at generating comprehensive and accurate code review comments and refining patches. We posit that this work would shed light on making informed decisions between MA- and SA-based approaches for complex software engineering tasks and challenge the trend of always adopting MA-based approaches to improve task performance. Future research should analyze the trade-off between MA- and SA-based approaches on other software engineering tasks. 

\section*{Limitations}

First, the evaluation of RCG and CR heavily depends on the LLM-as-a-judge, which may lead to a biased or hallucinated evaluation. While we followed the best practices and adopted the LLM-as-a-judge from the literature that has been proven to correlate with human experts, potential bias and inaccuracies may still exist in the evaluation results, especially for CR. In the future, we will try to curate a benchmark containing test cases to more accurately evaluate the quality of the refined patches. Second, our evaluation results may not be readily generalized to code patches written in programming languages other than Java. Similarly, the findings may not be fully transferable to all MA and SA architectures in the literature, and the backbone LLMs. Future work can incorporate more LLMs and a diverse range of MA/SA architectures. Third, the tools provided to the LLM in the ReAct settings may have internal limitations that can affect agents' performance. Our results may not be applicable to ReAct agents with a different set of tools. Moreover, each agent in \textit{MA-ReAct} is only equipped with one specialized tool for a single dimension, while \textit{SA-ReAct} agents can access all available tools. While expertise is distributed, doing so may hinder \textit{MA-ReAct} in cases where cross-dimensional tool synergy is important.  Finally, alternative prompting strategies and writing styles may produce different results.

% llm judge 
% implementation mistakes
% generalizability of results: evaluation dataset, MA architectures, issue dimensions and tools, the selected LLM, prompts, 

% This document does not cover the content requirements for ACL or any
% other specific venue.  Check the author instructions for
% information on
% maximum page lengths, the required ``Limitations'' section,
% and so on.

\section*{Ethical considerations}

We adopted \textit{CuRev} dataset \cite{sghaier2025harnessing} that is publicly available and has been used in prior works. We do not use any proprietary or confidential data. We strictly adhered to the dataset's licensing terms, intended usage, and ethical guidelines. For LLM usage, we used their public APIs and followed their specified functionalities and licensing agreements, ensuring alignment with ethical norms and policies. Since our work studies code review automation techniques, we encourage developers to use them as a source of information with appropriate human oversight. The reason lies in that these techniques may still generate misleading review comments or defective code patches that would lead to unintended software defects and vulnerabilities.

% \section*{Acknowledgments}

% We appreciate the insights and fruitful discussions with Iftekhar Ahmed and Hoa Khanh Dam. This work was supported by the computational resources and services provided by the Informatics department at the University of California, Irvine. We thank anonymous reviewers for their invaluable feedback.

% This document has been adapted
% by Steven Bethard, Ryan Cotterell and Rui Yan
% from the instructions for earlier ACL and NAACL proceedings, including those for
% ACL 2019 by Douwe Kiela and Ivan Vuli\'{c},
% NAACL 2019 by Stephanie Lukin and Alla Roskovskaya,
% ACL 2018 by Shay Cohen, Kevin Gimpel, and Wei Lu,
% NAACL 2018 by Margaret Mitchell and Stephanie Lukin,
% Bib\TeX{} suggestions for (NA)ACL 2017/2018 from Jason Eisner,
% ACL 2017 by Dan Gildea and Min-Yen Kan,
% NAACL 2017 by Margaret Mitchell,
% ACL 2012 by Maggie Li and Michael White,
% ACL 2010 by Jing-Shin Chang and Philipp Koehn,
% ACL 2008 by Johanna D. Moore, Simone Teufel, James Allan, and Sadaoki Furui,
% ACL 2005 by Hwee Tou Ng and Kemal Oflazer,
% ACL 2002 by Eugene Charniak and Dekang Lin,
% and earlier ACL and EACL formats written by several people, including
% John Chen, Henry S. Thompson and Donald Walker.
% Additional elements were taken from the formatting instructions of the \emph{International Joint Conference on Artificial Intelligence} and the \emph{Conference on Computer Vision and Pattern Recognition}.

\bibliography{custom}

\appendix

\section{Prompt Details}
\label{sec:appendix-prompts}

\begin{figure*}[!htbp]
  \includegraphics[width=\textwidth]{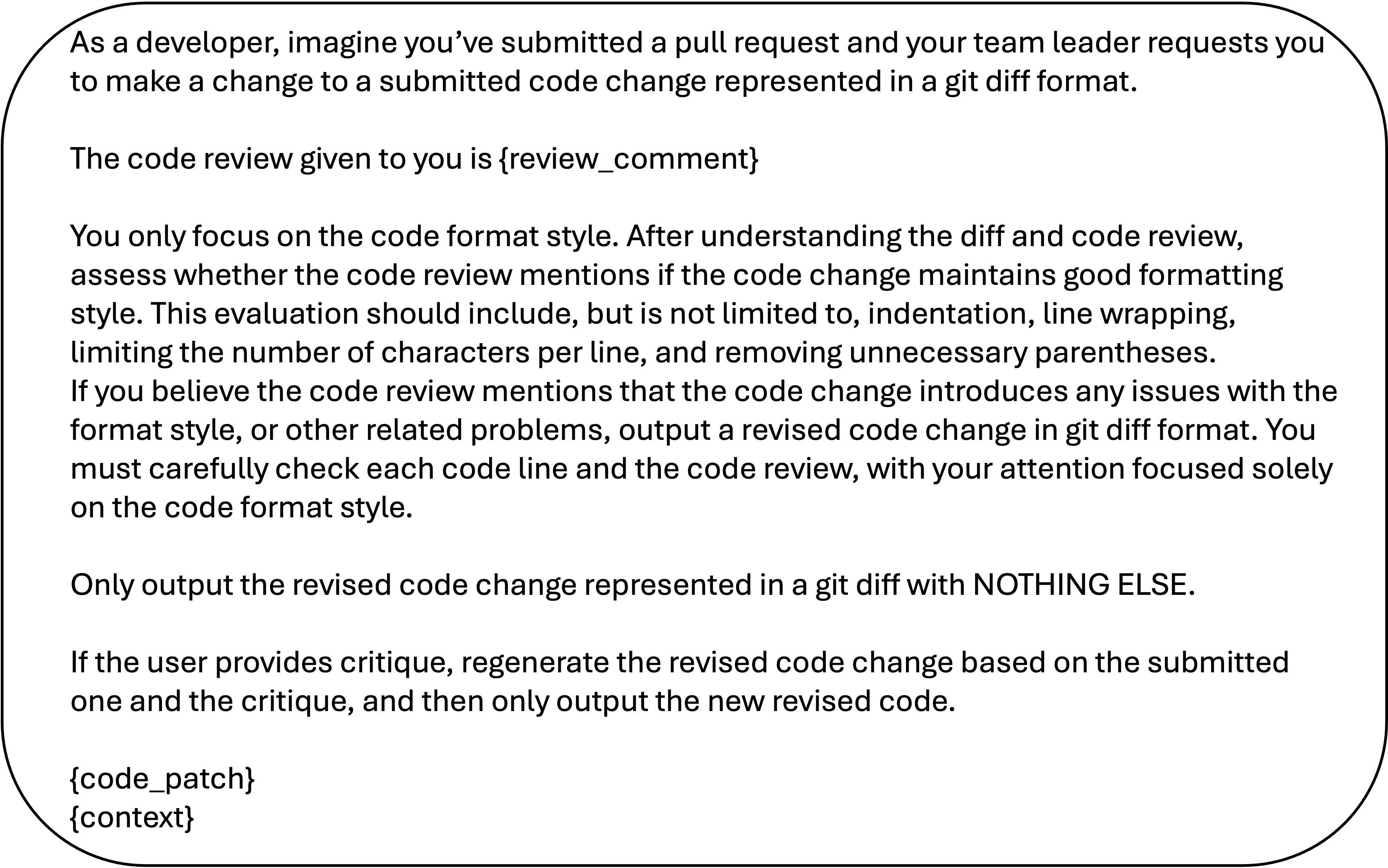}
  \caption{Adapted agent prompt (code formatting style) of \textit{Hydra Reviewer} on CR based on human-authored code review comments}
  \label{fig:hydra-cr-prompt}
\end{figure*}

\begin{figure*}[!htbp]
  \includegraphics[width=\textwidth]{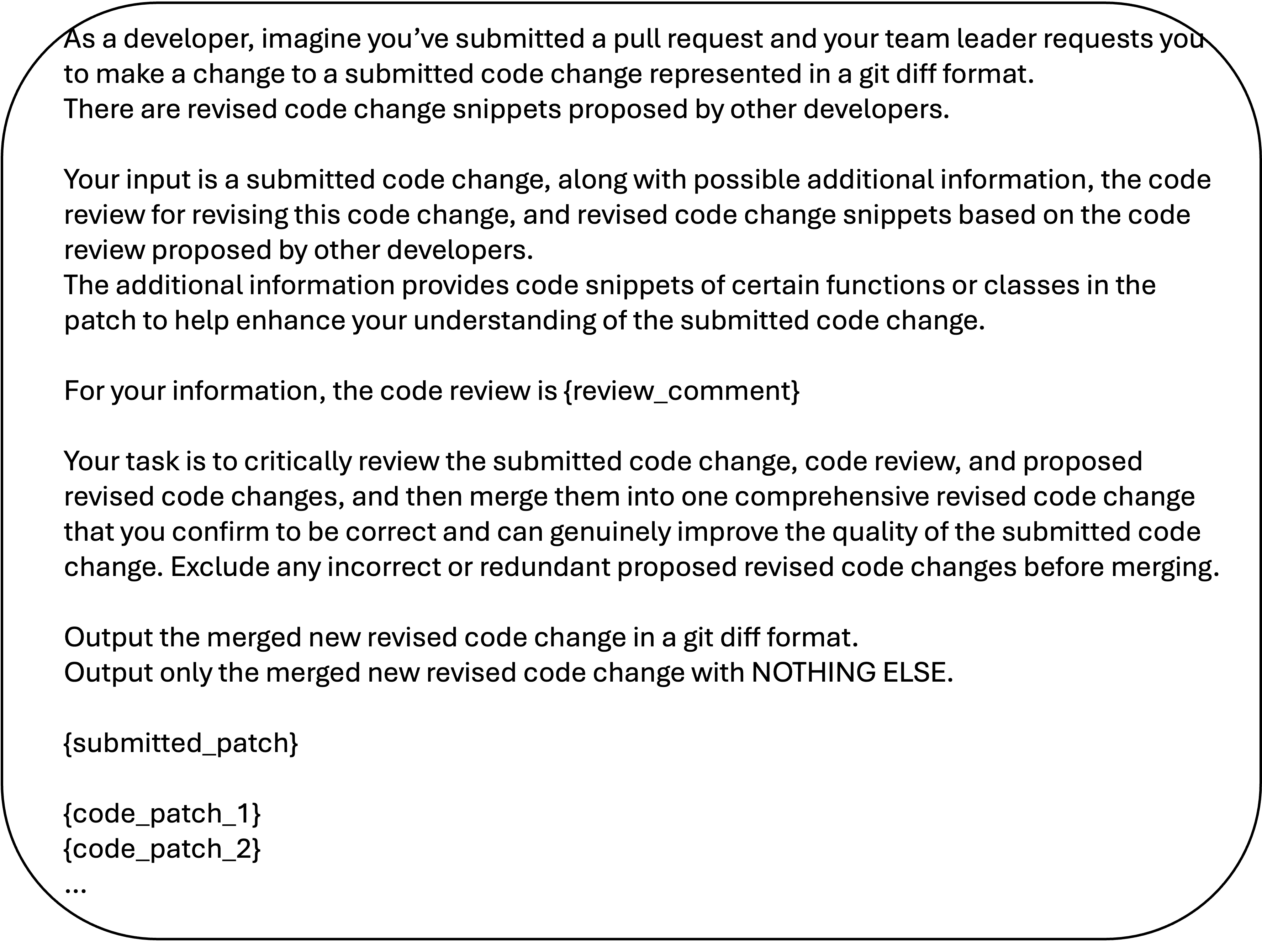}
  \caption{Adapted agent prompt (summarizer) of \textit{Hydra Reviewer} on CR based on human-authored code review comments}
  \label{fig:hydra-sum-prompt}
\end{figure*}

\begin{figure*}[!htbp]
  \includegraphics[width=\textwidth]{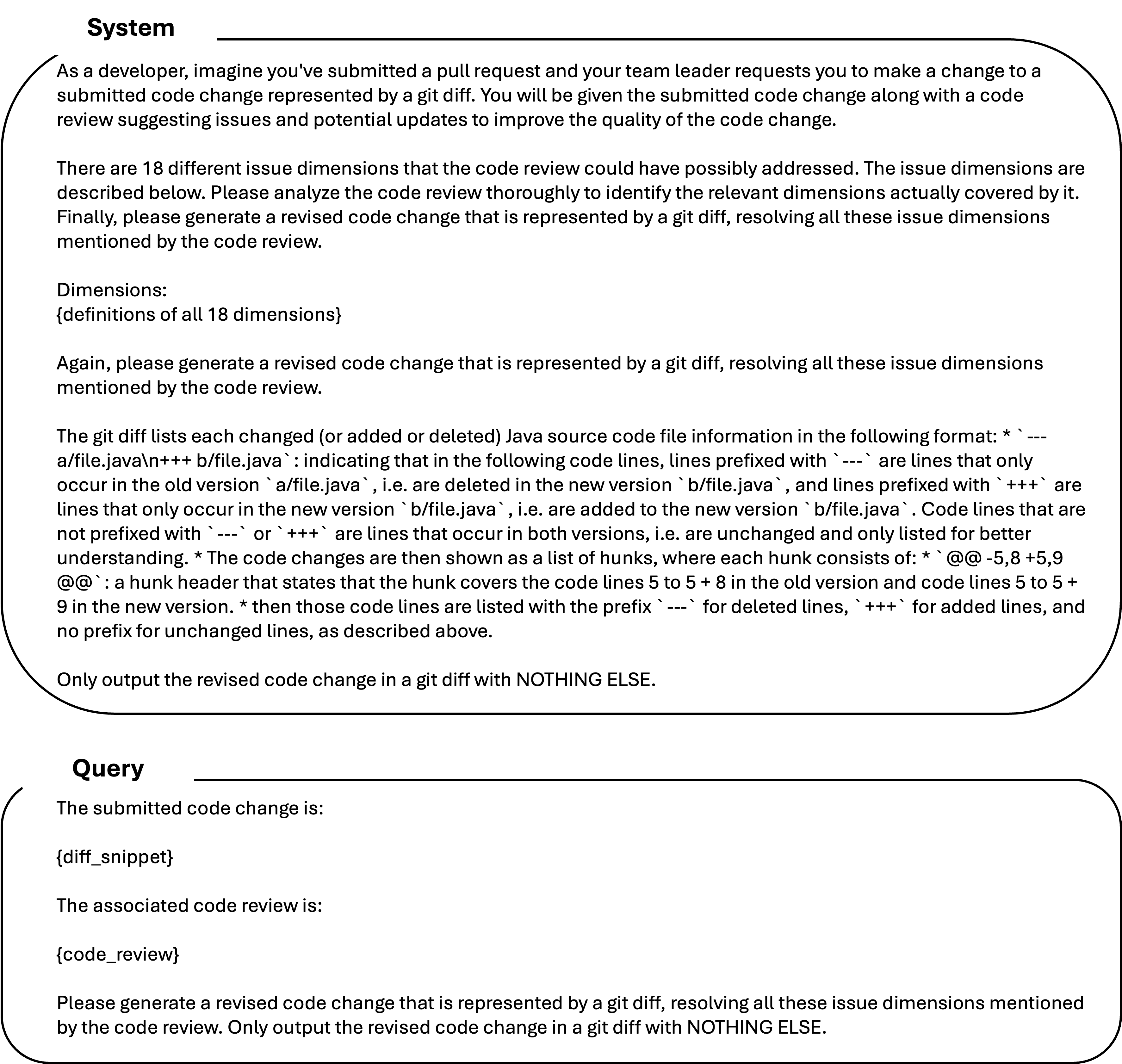}
  \caption{Adapted agent prompt of \textit{C-ChatGPT} on CR based on human-authored code review comments}
  \label{fig:c-chatgpt-cr}
\end{figure*}

\begin{figure*}[!htbp]
  \includegraphics[width=\textwidth]{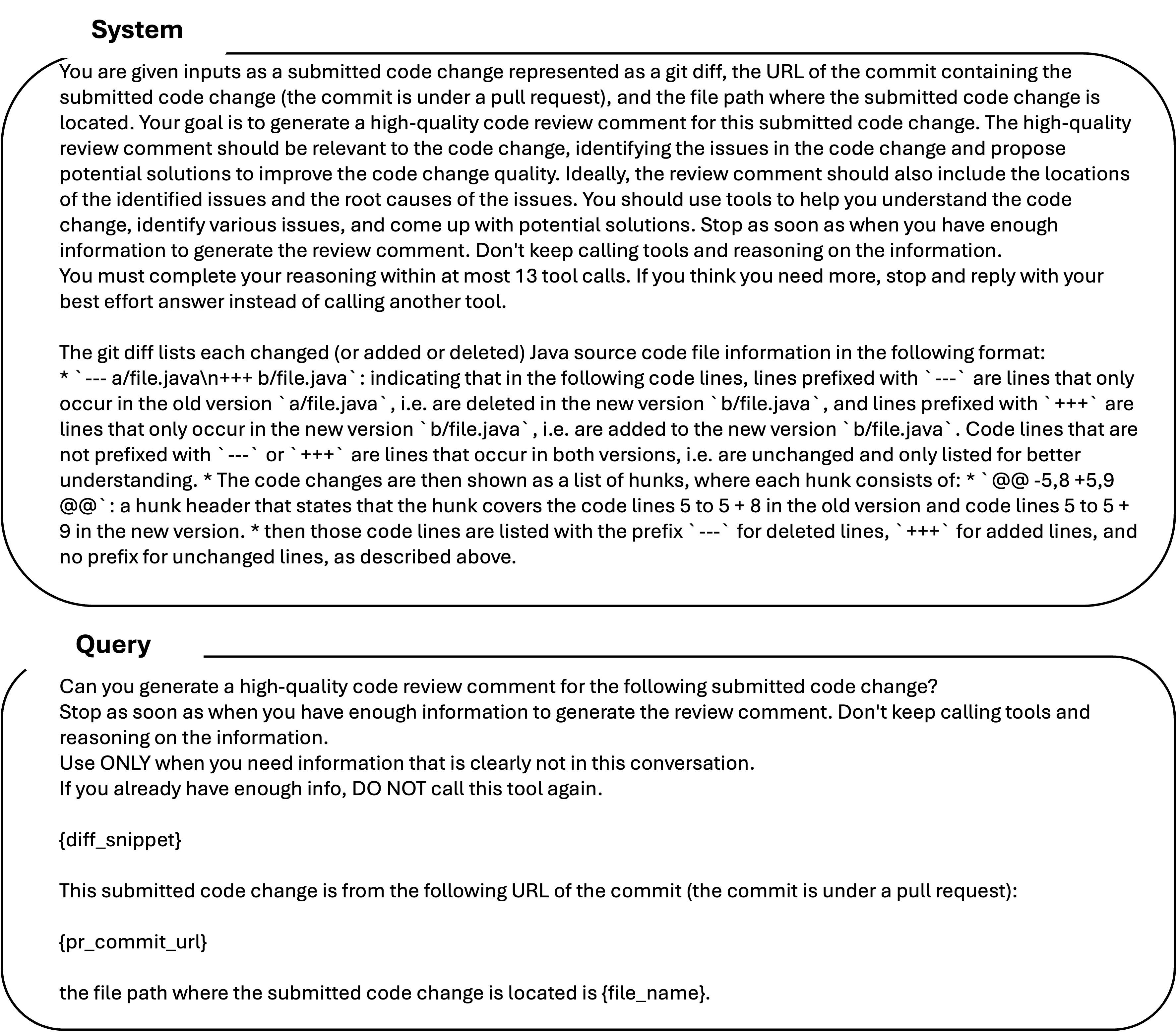}
  \caption{The prompt of \textit{SA-ReAct} on RCG}
  \label{fig:sa-react-rcg}
\end{figure*}

\begin{figure*}[!htbp]
  \includegraphics[width=\textwidth]{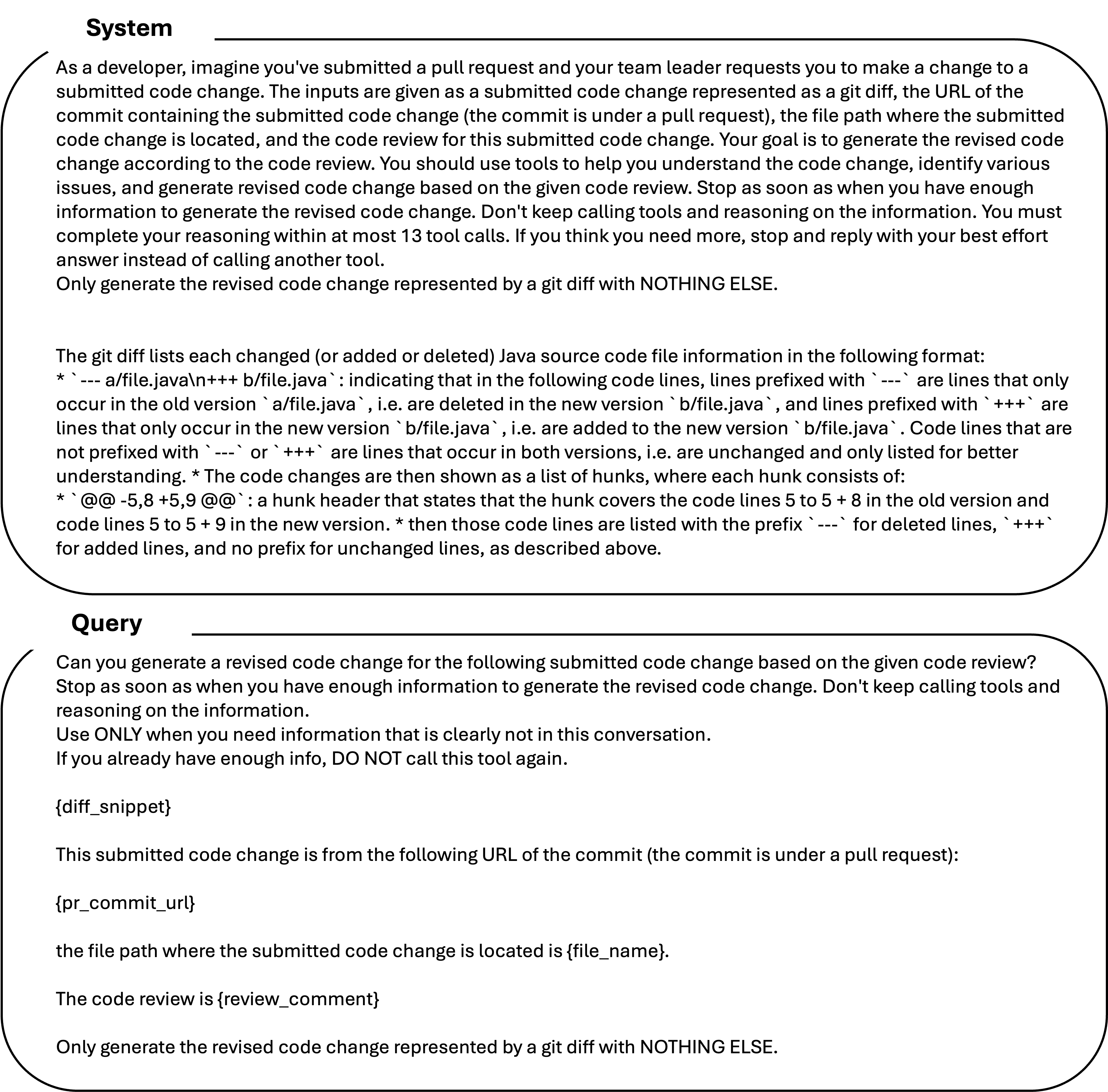}
  \caption{The prompt of \textit{SA-ReAct} on CR based on human-authored code review comments}
  \label{fig:sa-react-cr}
\end{figure*}

\begin{figure*}[!htbp]
  \includegraphics[width=\textwidth]{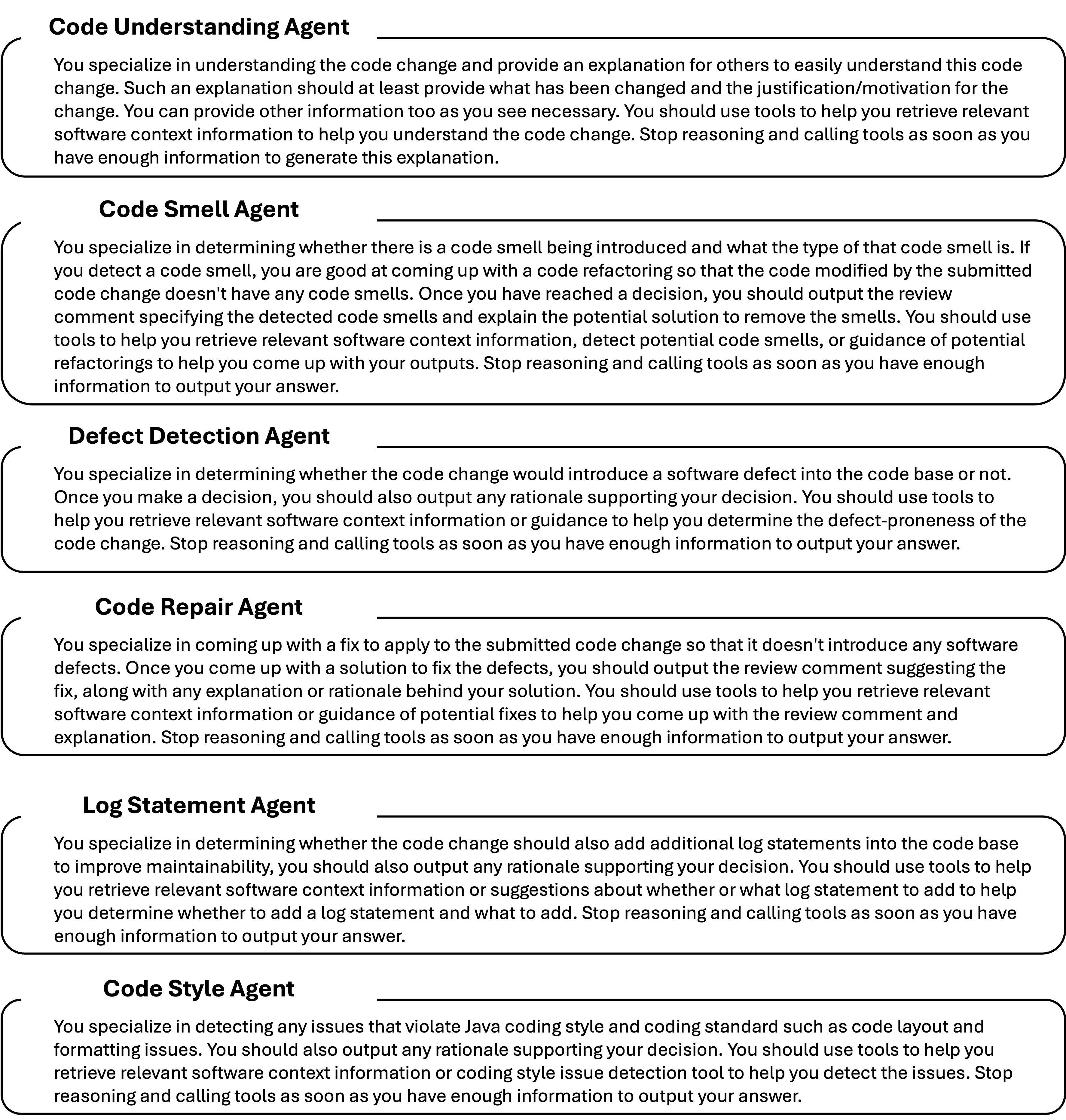}
  \caption{The system prompts of \textit{MA-ReAct} agents on RCG}
  \label{fig:ma-react-rcg}
\end{figure*}

\begin{figure*}[!htbp]
  \includegraphics[width=\textwidth]{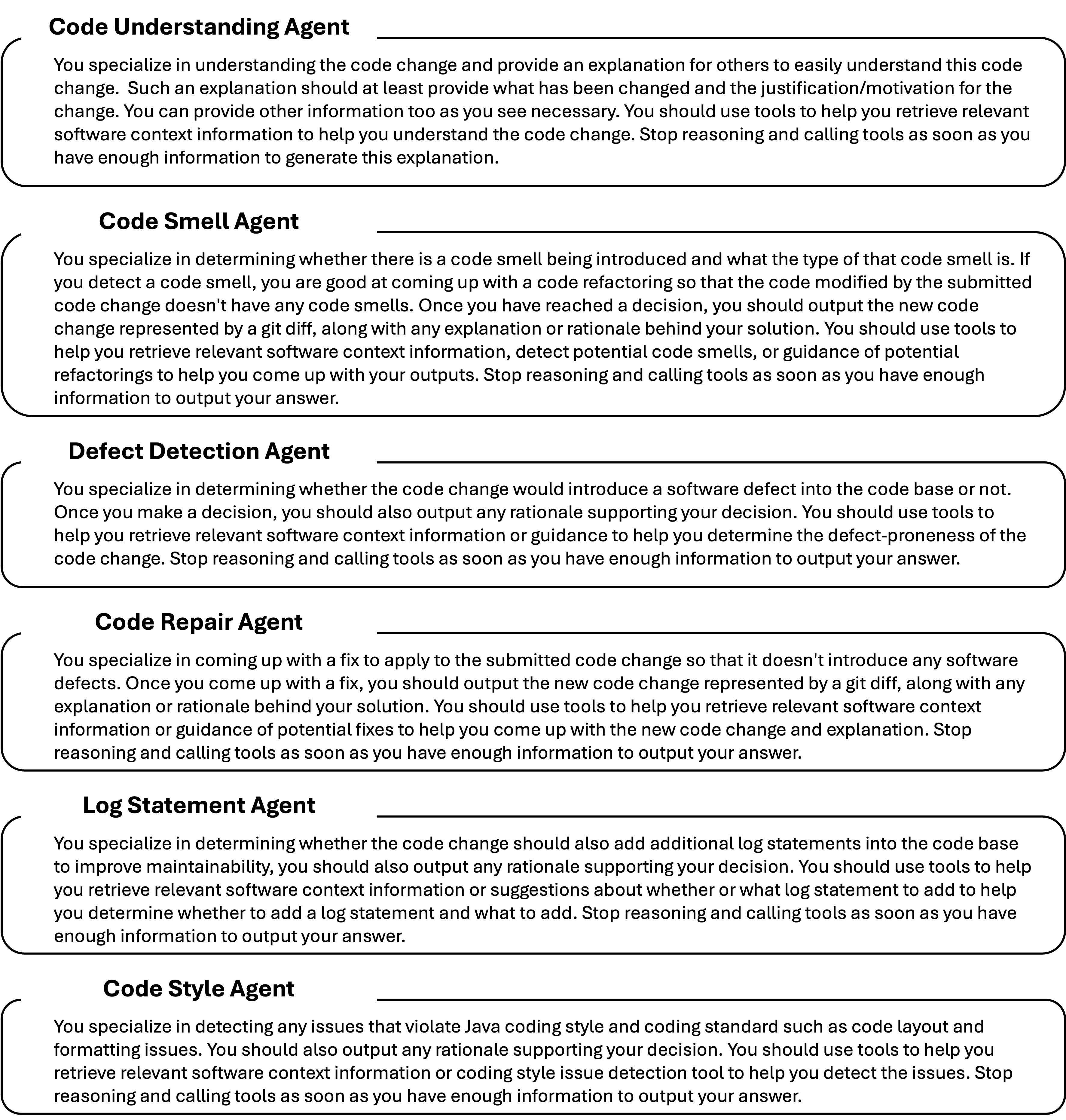}
  \caption{The system prompts of \textit{MA-ReAct} agents on CR based on human-authored code review comments}
  \label{fig:ma-react-cr}
\end{figure*}

\begin{figure*}[!htbp]
  \includegraphics[width=\textwidth]{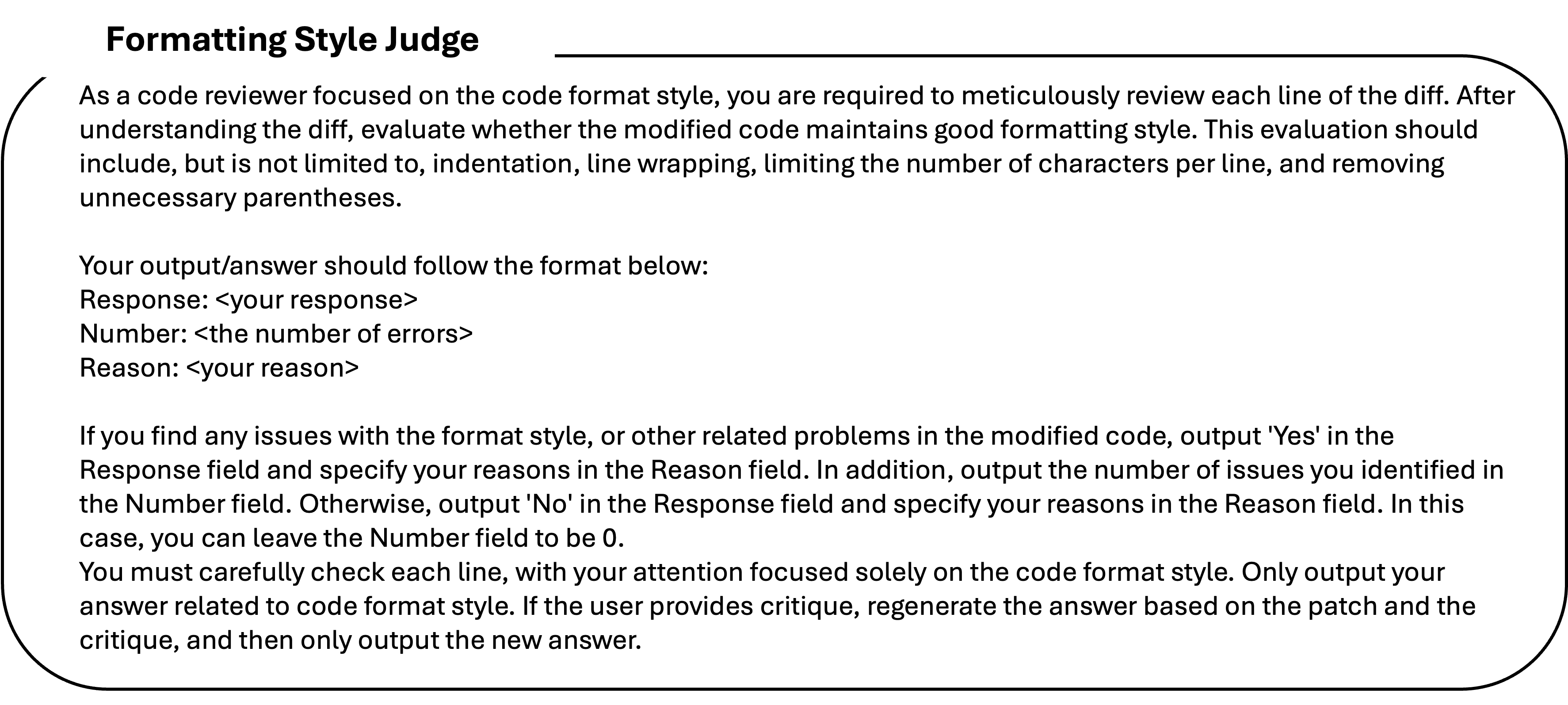}
  \caption{The LLM-as-a-judge for Code Quality evaluation (code formatting style)}
  \label{fig:judge}
\end{figure*}

The prompt of the adapted \textit{Hydra Reviewer} that automates CR by generating refined patches given the human-authored review comments is shown in Figure \ref{fig:hydra-cr-prompt}. It's worth noting that we only show the prompt for one agent out of all 18 specialized agents as the adaptation is the same across all the agents \cite{ren2025hydra}. 

The prompt of the adapted agent summarizing patches from various specialized agents in \textit{Hydra Reviewer} is shown in Figure \ref{fig:hydra-sum-prompt}. The prompt of the adapted \textit{C-ChatGPT} for refining patches based on human-authored review comments is shown in Figure \ref{fig:c-chatgpt-cr}. Brief descriptions of the 18 dimensions are presented in Table \ref{tab:18dim}. For more information about the complete agent design, components, and quality dimensions, refer to \cite{ren2025hydra}. 

We include the system and query prompts of \textit{SA-ReAct} agents in Figure \ref{fig:sa-react-cr} and \ref{fig:sa-react-rcg}. We implemented the \textit{SA-ReAct} agents using LangChain \cite{langchain}.

In Figure \ref{fig:ma-react-rcg}, we present the system prompts of all the specialized agents in the \textit{MA-ReAct} on RCG. Figure \ref{fig:ma-react-cr} shows their system prompts on CR based on human-authored review comments. We implemented the MA architecture of \textit{MA-ReAct} using LangGraph \cite{langgraph}. It's worth noting that the query prompts to the \textit{MA-ReAct} remain consistent with what's presented in Figure \ref{fig:sa-react-rcg} and \ref{fig:sa-react-cr}. Moreover, the difference between the agents on RCG and those on CR only lies in the instructions and task goals of their prompts. The tools that are equipped by the agents stay the same. Thus, we provide the detailed mapping between the tools in Table \ref{tab:patchund} and \ref{tab:qualityassist} and the agents of \textit{MA-ReAct} in Table \ref{tab:mareact-tools}.

We provide the prompt of the LLM-as-a-judge for evaluating Code Quality in Figure \ref{fig:judge}. We only provide the prompt of the LLM-as-a-judge in evaluating code formatting style dimension, given that the only difference among the agents is the description of the issue dimension. For detailed issue dimension descriptions of all 18 dimensions, refer to \cite{ren2025hydra}.

% \usepackage{multirow}
\begin{table*}[!htbp]
\centering
\caption{Brief descriptions of the 18 dimensions by \textit{Hydra Reviewer}}
\label{tab:18dim}
\resizebox{\textwidth}{!}{%
\begin{tabular}{l|cccccc}
\hline
                                      & \textbf{Formatting Style}                                                                                                                                   & \textbf{Logic Readability}                                                                                                     & \textbf{Semantic Correctness}                                                                                                                            & \textbf{Syntax Correctness}                                                                                                                  & \textbf{Testing}                                                                                                                                    & \textbf{Comment Quality}                                                                                                           \\ \hline
\multirow{2}{*}{\textbf{Description}} & \multirow{2}{*}{\begin{tabular}[c]{@{}c@{}}whether the code format follows the project's \\ formatting style\end{tabular}} & \multirow{2}{*}{\begin{tabular}[c]{@{}c@{}}how easy the code logic is\\ to understand\end{tabular}}                            & \multirow{2}{*}{\begin{tabular}[c]{@{}c@{}}whether code semantics is not properly\\ implemented as required\end{tabular}}                                & \multirow{2}{*}{\begin{tabular}[c]{@{}c@{}}whether there are code\\ syntax issues\end{tabular}}                                              & \multirow{2}{*}{\begin{tabular}[c]{@{}c@{}}whether tests cover\\ the new patch\end{tabular}}                                                        & \multirow{2}{*}{\begin{tabular}[c]{@{}c@{}}how readable or outdated\\ are the comments\end{tabular}}                               \\
                                      &                                                                                                                                                             &                                                                                                                                &                                                                                                                                                          &                                                                                                                                              &                                                                                                                                                     &                                                                                                                                    \\ \hline
                                      & \textbf{Comment Style}                                                                                                                                      & \textbf{Compatibility}                                                                                                         & \textbf{Extensibility}                                                                                                                                   & \textbf{Fault Tolerance}                                                                                                                     & \textbf{Identifier Naming Readability}                                                                                                              & \textbf{Identifier Naming Style}                                                                                                   \\ \hline
\multirow{2}{*}{\textbf{Description}} & \multirow{2}{*}{\begin{tabular}[c]{@{}c@{}}whether the comments are consistent \\ with project's comment style\end{tabular}}                                & \multirow{2}{*}{\begin{tabular}[c]{@{}c@{}}how compatible is the patch\\ across different versions and platforms\end{tabular}} & \multirow{2}{*}{\begin{tabular}[c]{@{}c@{}}whether the patch has good extensibility \\ to accommodate potential future requirement changes\end{tabular}} & \multirow{2}{*}{\begin{tabular}[c]{@{}c@{}}how good can the code handle unexpected\\ inputs, whether there is validation check\end{tabular}} & \multirow{2}{*}{\begin{tabular}[c]{@{}c@{}}how clear the identifiers are\\ in the patch\end{tabular}}                                               & \multirow{2}{*}{\begin{tabular}[c]{@{}c@{}}whether the identifiers in the patch\\  follow the project’s naming style\end{tabular}} \\
                                      &                                                                                                                                                             &                                                                                                                                &                                                                                                                                                          &                                                                                                                                              &                                                                                                                                                     &                                                                                                                                    \\ \hline
                                      & \textbf{Name and Logic Consistency}                                                                                                                         & \textbf{Performance}                                                                                                           & \textbf{Programming Handling Conventions}                                                                                                                & \textbf{Redundancy}                                                                                                                          & \textbf{Runtime Observability}                                                                                                                      & \textbf{Security Compliance}                                                                                                       \\ \hline
\multirow{2}{*}{\textbf{Description}} & \multirow{2}{*}{\begin{tabular}[c]{@{}c@{}}how consistent between the patch and\\ the code identifiers\end{tabular}}                                        & \multirow{2}{*}{\begin{tabular}[c]{@{}c@{}}the resource consumption, time and\\ space complexity of the code\end{tabular}}     & \multirow{2}{*}{\begin{tabular}[c]{@{}c@{}}whether there is any failure following project-specific \\ programming conventions in the patch\end{tabular}} & \multirow{2}{*}{\begin{tabular}[c]{@{}c@{}}whether there is duplicateor redundant code\\ blocks or logic in the patch\end{tabular}}          & \multirow{2}{*}{\begin{tabular}[c]{@{}c@{}}whether there should be changes to logging, \\ assertions in the patch that trace behavior\end{tabular}} & \multirow{2}{*}{\begin{tabular}[c]{@{}c@{}}whether the patch can introduce\\ security vulnerabilities\end{tabular}}                \\
                                      &                                                                                                                                                             &                                                                                                                                &                                                                                                                                                          &                                                                                                                                              &                                                                                                                                                     &                                                                                                                                    \\ \hline
\end{tabular}
}
\end{table*}

\begin{table*}[!htbp]
\centering
\caption{The ReAct agents of \textit{MA-ReAct} and the tools equipped by them}
\label{tab:mareact-tools}
\resizebox{0.8\textwidth}{!}{%
\begin{tabular}{l|l}
\hline
\textbf{Agent}          & \textbf{Equipped Tools}                                                                                 \\ \hline
Code Understanding Agent & All \textit{patch understanding tools} in Table \ref{tab:patchund}                                                                 \\ \hline
Code Smell Agent         & \begin{tabular}[c]{@{}l@{}}All \textit{patch understanding tools} in Table \ref{tab:patchund}\\ DesigniteJava\\ iSMELL\end{tabular} \\ \hline
Defect Detection Agent   & \begin{tabular}[c]{@{}l@{}}All \textit{patch understanding tools} in Table \ref{tab:patchund}\\ LApredict\end{tabular}              \\ \hline
Code Repair Agent        & \begin{tabular}[c]{@{}l@{}}All \textit{patch understanding tools} in Table \ref{tab:patchund}\\ D4C\end{tabular}                    \\ \hline
Log Statement Agent      & \begin{tabular}[c]{@{}l@{}}All \textit{patch understanding tools} in Table \ref{tab:patchund}\\ FastLog\end{tabular}                \\ \hline
Code Style Agent         & \begin{tabular}[c]{@{}l@{}}All \textit{patch understanding tools} in Table \ref{tab:patchund}\\ Checkstyle\end{tabular}             \\ \hline
\end{tabular}
}
\end{table*}

\section{Patch Understanding Tools}
\label{sec:appendix-patchundtools}

We included a wide range of tools to retrieve software contexts associated with the submitted patch in Table \ref{tab:patchund}. The goal is to help ReAct agents obtain a comprehensive overview of the submitted patches to generate high-quality review comments/refined patches. These tools have been shown to be useful and effective for enhancing code patch understanding. For more information about these tools, refer to \cite{li2024only, li2025consider}.

\begin{table*}[!htbp]
\caption{The descriptions of the \textit{patch understanding} tools for the ReAct agents}
\label{tab:patchund}
\resizebox{\textwidth}{!}{%
\begin{tabular}{l|l}
\hline
\textbf{Tool}                  & \textbf{Functionality}                                                                                                                                                                                            \\ \hline
Important File Information     & \begin{tabular}[c]{@{}l@{}}Identify the file that most other files in the project depend on. A static\\ analysis tool, Understand \cite{understand}, is used to identify file dependencies.\end{tabular}                            \\ \hline
Commit Type Information        & \begin{tabular}[c]{@{}l@{}}Identify the types of software maintenance activities of the patch. The\\ types include fixing software bugs, improving system design, and intro-\\ -ducing new features.\end{tabular} \\ \hline
Pull Request/Issue Report      & \begin{tabular}[c]{@{}l@{}}Retrieve the associated pull request and issue report contents, since they\\ provide the motivation/reason behind the patch.\end{tabular}                                              \\ \hline
Method Body Summary            & Generate the summary of the method that the patch is located in.                                                                                                                                                  \\ \hline
Class Body Summary             & Generate the summary of the class that the patch is located in.                                                                                                                                                   \\ \hline
Complete Enclosing Code Blocks & Obtain the smallest enclosing statement block that contains the patch.                                                                                                                                            \\ \hline
Excluded Callee Knowledge      & Generate the summaries of the methods invoked in the patch.                                                                                                                                                       \\ \hline
Excluded Variable Data Types   & \begin{tabular}[c]{@{}l@{}}Identify the data types and class information of variables referenced \\ in the patch.\end{tabular}                                                                                    \\ \hline
\end{tabular}
}
\end{table*}

\section{Quality Dimension Assistance Tools}
\label{sec:appendix-qualitytools}

We present the quality dimension assistance tools for the ReAct agents in Table \ref{tab:qualityassist}. These tools detect various software quality issues and propose solutions, thereby making the agents' responses on RCG and CR more reliable and factually grounded. For more details about their design and implementation, refer to the corresponding reference in Table \ref{tab:qualityassist}.

\begin{table*}[!htbp]
\caption{The descriptions of the \textit{quality-dimension assistance} tools for the ReAct agents}
\label{tab:qualityassist}
\resizebox{\textwidth}{!}{%
\begin{tabular}{l|l|l}
\hline
\textbf{Dimension}    & \textbf{Tool} & \textbf{Functionality}                                                                                                                                      \\ \hline
Code Smell Detection  & DesigniteJava \cite{Sharma2016} & \begin{tabular}[c]{@{}l@{}}Detect various code smells including architecture smells, design smells,\\ implementation smells, and test smells.\end{tabular}  \\ \hline
Code Refactoring      & iSMELL \cite{wu2024ismell}       & Refactor the identified code smells in the submitted code patch.                                                                                            \\ \hline
Defect Prediction     & LApredict \cite{zeng2021deep}    & \begin{tabular}[c]{@{}l@{}}Predict whether a given code patch can introduce software defects into\\ the code base.\end{tabular}                             \\ \hline
Code Repair           & D4C \cite{xu2025aligning}          & \begin{tabular}[c]{@{}l@{}}Refine code and eliminate software defects without first identifying \\ faulty statements.\end{tabular}                          \\ \hline
Log Statements        & FastLog \cite{xie2024fastlog}      & \begin{tabular}[c]{@{}l@{}}Predict the insertion position of needed log statements, and then genera-\\ -te a complete log statement to insert.\end{tabular} \\ \hline
Code Formatting Style & Checkstyle \cite{checkstyle}   & \begin{tabular}[c]{@{}l@{}}Detect violations of a coding standard, including design problems and \\ layout/formatting issues.\end{tabular}                  \\ \hline
\end{tabular}
}
\end{table*}

\section{Detailed Evaluation Results}
\label{sec:appendix-evalresults}

% \usepackage{multirow}
\begin{table*}[!htbp]
\centering
\caption{The statistical tests of the evaluation results of all techniques for RCG. We conducted statistical tests on the score-based metrics, Conciseness, Clarity, and Relevance. We reported the p-value of the Welch's t-test and Cohen's D. A grey cell indicates the difference is significant.}
\label{tab:stat-test-rcg}
\resizebox{0.8\textwidth}{!}{%
\begin{tabular}{l|ccc}
\hline
\textbf{}               & \textbf{Conciseness}                                                                                                  & \textbf{Clarity}                                                                                                      & \textbf{Relevance}                                                                                                 \\ \hline
\textbf{Hydra Reviewer} & \cellcolor[HTML]{C0C0C0}                                                                                              & \cellcolor[HTML]{C0C0C0}                                                                                              & \cellcolor[HTML]{C0C0C0}                                                                                           \\
\textbf{C-ChatGPT}      & \multirow{-2}{*}{\cellcolor[HTML]{C0C0C0}\begin{tabular}[c]{@{}c@{}}p-value=1.853e-21\\ Cohen's D=0.992\end{tabular}} & \multirow{-2}{*}{\cellcolor[HTML]{C0C0C0}\begin{tabular}[c]{@{}c@{}}p-value=6.669e-12\\ Cohen's D=0.699\end{tabular}} & \multirow{-2}{*}{\cellcolor[HTML]{C0C0C0}\begin{tabular}[c]{@{}c@{}}p-value=0.0006\\ Cohen's D=0.345\end{tabular}} \\ \hline
\textbf{MA-ReAct}       & \cellcolor[HTML]{C0C0C0}                                                                                              & \cellcolor[HTML]{C0C0C0}                                                                                              & \cellcolor[HTML]{C0C0C0}                                                                                           \\
\textbf{SA-ReAct}       & \multirow{-2}{*}{\cellcolor[HTML]{C0C0C0}\begin{tabular}[c]{@{}c@{}}p-value=0.003\\ Cohen's D=0.237\end{tabular}}     & \multirow{-2}{*}{\cellcolor[HTML]{C0C0C0}\begin{tabular}[c]{@{}c@{}}p-value=1.165e-8\\ Cohen's D=0.458\end{tabular}}  & \multirow{-2}{*}{\cellcolor[HTML]{C0C0C0}\begin{tabular}[c]{@{}c@{}}p-value=0.0002\\ Cohen's D=0.298\end{tabular}} \\ \hline
\end{tabular}
}
\end{table*}

To analyze the statistical significance of the score differences in Section \ref{sec:results}, we conducted Welch's t-test \cite{ruxton2006unequal} to determine whether the scores by different techniques (MA and its SA counterpart) are significantly different when the two score groups may have different variances. We also performed Cohen's D \cite{diener2010cohen} to measure the effect size. 

Table \ref{tab:stat-test-rcg} shows the results of the statistical analysis on the score-based metrics used to evaluate the techniques on RCG, demonstrating all score differences between MA and their SA counterparts are statistically significant with effect sizes ranging from small (0.237) to large (0.992).

Table \ref{tab:stat-hydra-chatgptcr} and \ref{tab:stat-hydra-hm} report the statistical tests of the Code Quality (Section \ref{sec:CR_eval_metrics}) evaluation results of \textit{Hydra Reviewer} and \textit{C-ChatGPT} on the two settings to perform CR. Statistical test and effect size analysis are conducted on the values of two techniques, MA and SA, in a quality dimension (i.e., number of issues in that dimension). We report the results across all dimensions. Table \ref{tab:stat-react-chatgptcr} and \ref{tab:stat-react-hm} present those of \textit{MA-ReAct} and \textit{SA-ReAct}. In addition, we show the average values of the Code Quality evaluation for \textit{MA-ReAct} and \textit{SA-ReAct} in Table \ref{tab:llmeval_react_humancomment} and \ref{tab:llmeval_react_generatedcomment}.

% \usepackage{multirow}
\begin{table*}[!htbp]
\caption{The code quality evaluation results of \textit{Hydra Reviewer} and \textit{C-ChatGPT} for CR. The refined patches are generated by ChatGPT-CR based on comments from different techniques. We report the statistical tests (Welch's t-test and Cohen's D) to analyze the number differences of quality issues detected by the LLM-as-a-judge. A grey cell indicates the value difference is statistically significant.}
\label{tab:stat-hydra-chatgptcr}
\resizebox{\textwidth}{!}{%
\begin{tabular}{l|cccccc}
\hline
                        & \textbf{Formatting Style}                                                                                                                 & \textbf{Logic Readability}                                                                                           & \textbf{Semantic Correctness}                                                                                        & \textbf{Syntax Correctness}                                                                                          & \textbf{Testing}                                                                                                      & \textbf{Comment Quality}                                                                                              \\ \hline
\textbf{Hydra Reviewer} & \cellcolor[HTML]{C0C0C0}{\color[HTML]{000000} }                                                                                           & \cellcolor[HTML]{C0C0C0}                                                                                             & \cellcolor[HTML]{C0C0C0}                                                                                             &                                                                                                                      & \cellcolor[HTML]{C0C0C0}                                                                                              & \cellcolor[HTML]{C0C0C0}                                                                                              \\
\textbf{C-ChatGPT}      & \multirow{-2}{*}{\cellcolor[HTML]{C0C0C0}{\color[HTML]{000000} \begin{tabular}[c]{@{}c@{}}p-value=0.0001\\ Cohen's D=0.302\end{tabular}}} & \multirow{-2}{*}{\cellcolor[HTML]{C0C0C0}\begin{tabular}[c]{@{}c@{}}p-value=1.867e-8\\ Cohen's D=0.451\end{tabular}} & \multirow{-2}{*}{\cellcolor[HTML]{C0C0C0}\begin{tabular}[c]{@{}c@{}}p-value=0.0001\\ Cohen's D=0.304\end{tabular}}   & \multirow{-2}{*}{\begin{tabular}[c]{@{}c@{}}p-value=0.070\\ Cohen's D=0.144\end{tabular}}                            & \multirow{-2}{*}{\cellcolor[HTML]{C0C0C0}\begin{tabular}[c]{@{}c@{}}p-value=2.326e-13\\ Cohen's D=0.593\end{tabular}} & \multirow{-2}{*}{\cellcolor[HTML]{C0C0C0}\begin{tabular}[c]{@{}c@{}}p-value=1.475e-18\\ Cohen's D=0.718\end{tabular}} \\ \hline
                        & \textbf{Comment Style}                                                                                                                    & \textbf{Compatibility}                                                                                               & \textbf{Extensibility}                                                                                               & \textbf{Fault Tolerance}                                                                                             & \textbf{Identifier Naming Readability}                                                                                & \textbf{Identifier Naming Style}                                                                                      \\ \hline
\textbf{Hydra Reviewer} & \cellcolor[HTML]{C0C0C0}                                                                                                                  & \cellcolor[HTML]{C0C0C0}                                                                                             & \cellcolor[HTML]{C0C0C0}                                                                                             &                                                                                                                      & \cellcolor[HTML]{C0C0C0}                                                                                              & \cellcolor[HTML]{C0C0C0}                                                                                              \\
\textbf{C-ChatGPT}      & \multirow{-2}{*}{\cellcolor[HTML]{C0C0C0}\begin{tabular}[c]{@{}c@{}}p-value=3.344e-17\\ Cohen's D=0.687\end{tabular}}                     & \multirow{-2}{*}{\cellcolor[HTML]{C0C0C0}\begin{tabular}[c]{@{}c@{}}p-value=1.110e-8\\ Cohen's D=0.458\end{tabular}} & \multirow{-2}{*}{\cellcolor[HTML]{C0C0C0}\begin{tabular}[c]{@{}c@{}}p-value=3.909e-9\\ Cohen's D=0.473\end{tabular}} & \multirow{-2}{*}{\begin{tabular}[c]{@{}c@{}}p-value=0.517\\ Cohen's D=0.051\end{tabular}}                            & \multirow{-2}{*}{\cellcolor[HTML]{C0C0C0}\begin{tabular}[c]{@{}c@{}}p-value=4.479e-6\\ Cohen's D=0.366\end{tabular}}  & \multirow{-2}{*}{\cellcolor[HTML]{C0C0C0}\begin{tabular}[c]{@{}c@{}}p-value=3.245e-5\\ Cohen's D=0.331\end{tabular}}  \\ \hline
                        & \textbf{Name and Logic Consistency}                                                                                                       & \textbf{Performance}                                                                                                 & \textbf{Programming Handling Conventions}                                                                            & \textbf{Redundancy}                                                                                                  & \textbf{Runtime Observability}                                                                                        & \textbf{Security Compliance}                                                                                          \\ \hline
\textbf{Hydra Reviewer} & \cellcolor[HTML]{C0C0C0}                                                                                                                  & \cellcolor[HTML]{C0C0C0}                                                                                             & \cellcolor[HTML]{C0C0C0}                                                                                             & \cellcolor[HTML]{C0C0C0}                                                                                             & \cellcolor[HTML]{C0C0C0}                                                                                              &                                                                                                                       \\
\textbf{C-ChatGPT}      & \multirow{-2}{*}{\cellcolor[HTML]{C0C0C0}\begin{tabular}[c]{@{}c@{}}p-value=1.586e-8\\ Cohen's D=0.453\end{tabular}}                      & \multirow{-2}{*}{\cellcolor[HTML]{C0C0C0}\begin{tabular}[c]{@{}c@{}}p-value=9.341e-8\\ Cohen's D=0.428\end{tabular}} & \multirow{-2}{*}{\cellcolor[HTML]{C0C0C0}\begin{tabular}[c]{@{}c@{}}p-value=2.399e-9\\ Cohen's D=0.479\end{tabular}} & \multirow{-2}{*}{\cellcolor[HTML]{C0C0C0}\begin{tabular}[c]{@{}c@{}}p-value=1.181e-6\\ Cohen's D=0.388\end{tabular}} & \multirow{-2}{*}{\cellcolor[HTML]{C0C0C0}\begin{tabular}[c]{@{}c@{}}p-value=7.616e-10\\ Cohen's D=0.495\end{tabular}} & \multirow{-2}{*}{\begin{tabular}[c]{@{}c@{}}p-value=0.220\\ Cohen's D=0.097\end{tabular}}                             \\ \hline
\end{tabular}
}
\end{table*}

% \usepackage{multirow}
\begin{table*}[!htbp]
\caption{The code quality evaluation results of adapted \textit{Hydra Reviewer} and \textit{C-ChatGPT} for CR. The refined patches are generated by these adapted techniques based on comments from human reviewers. We report the statistical tests (Welch's t-test and Cohen's D) to analyze the number differences of quality issues detected by the LLM-as-a-judge. A grey cell indicates the value difference is statistically significant.}
\label{tab:stat-hydra-hm}
\resizebox{\textwidth}{!}{%
\begin{tabular}{l|cccccc}
\hline
                        & \textbf{Formatting Style}                                                                                                                   & \textbf{Logic Readability}                                                                                           & \textbf{Semantic Correctness}                                                                                        & \textbf{Syntax Correctness}                                                                                          & \textbf{Testing}                                                                                                     & \textbf{Comment Quality}                                                                                             \\ \hline
\textbf{Hydra Reviewer} & \cellcolor[HTML]{C0C0C0}{\color[HTML]{000000} }                                                                                             & \cellcolor[HTML]{C0C0C0}                                                                                             & \cellcolor[HTML]{FFFFFF}                                                                                             & \cellcolor[HTML]{C0C0C0}                                                                                             & \cellcolor[HTML]{C0C0C0}                                                                                             & \cellcolor[HTML]{C0C0C0}                                                                                             \\
\textbf{C-ChatGPT}      & \multirow{-2}{*}{\cellcolor[HTML]{C0C0C0}{\color[HTML]{000000} \begin{tabular}[c]{@{}c@{}}p-value=3.444e-7\\ Cohen's D=0.410\end{tabular}}} & \multirow{-2}{*}{\cellcolor[HTML]{C0C0C0}\begin{tabular}[c]{@{}c@{}}p-value=1.966e-5\\ Cohen's D=0.342\end{tabular}} & \multirow{-2}{*}{\cellcolor[HTML]{FFFFFF}\begin{tabular}[c]{@{}c@{}}p-value=0.263\\ Cohen's D=0.090\end{tabular}}    & \multirow{-2}{*}{\cellcolor[HTML]{C0C0C0}\begin{tabular}[c]{@{}c@{}}p-value=0.002\\ Cohen's D=0.246\end{tabular}}    & \multirow{-2}{*}{\cellcolor[HTML]{C0C0C0}\begin{tabular}[c]{@{}c@{}}p-value=2.610e-6\\ Cohen's D=0.377\end{tabular}} & \multirow{-2}{*}{\cellcolor[HTML]{C0C0C0}\begin{tabular}[c]{@{}c@{}}p-value=2.699e-5\\ Cohen's D=0.336\end{tabular}} \\ \hline
                        & \textbf{Comment Style}                                                                                                                      & \textbf{Compatibility}                                                                                               & \textbf{Extensibility}                                                                                               & \textbf{Fault Tolerance}                                                                                             & \textbf{Identifier Naming Readability}                                                                               & \textbf{Identifier Naming Style}                                                                                     \\ \hline
\textbf{Hydra Reviewer} & \cellcolor[HTML]{C0C0C0}                                                                                                                    & \cellcolor[HTML]{FFFFFF}                                                                                             & \cellcolor[HTML]{C0C0C0}                                                                                             & \cellcolor[HTML]{C0C0C0}                                                                                             & \cellcolor[HTML]{C0C0C0}                                                                                             & \cellcolor[HTML]{C0C0C0}                                                                                             \\
\textbf{C-ChatGPT}      & \multirow{-2}{*}{\cellcolor[HTML]{C0C0C0}\begin{tabular}[c]{@{}c@{}}p-value=4.839e-10\\ Cohen's D=0.503\end{tabular}}                       & \multirow{-2}{*}{\cellcolor[HTML]{FFFFFF}\begin{tabular}[c]{@{}c@{}}p-value=0.206\\ Cohen's D=0.101\end{tabular}}    & \multirow{-2}{*}{\cellcolor[HTML]{C0C0C0}\begin{tabular}[c]{@{}c@{}}p-value=0.0007\\ Cohen's D=0.271\end{tabular}}   & \multirow{-2}{*}{\cellcolor[HTML]{C0C0C0}\begin{tabular}[c]{@{}c@{}}p-value=0.017\\ Cohen's D=0.190\end{tabular}}    & \multirow{-2}{*}{\cellcolor[HTML]{C0C0C0}\begin{tabular}[c]{@{}c@{}}p-value=3.417e-5\\ Cohen's D=0.332\end{tabular}} & \multirow{-2}{*}{\cellcolor[HTML]{C0C0C0}\begin{tabular}[c]{@{}c@{}}p-value=1.566e-6\\ Cohen's D=0.385\end{tabular}} \\ \hline
                        & \textbf{Name and Logic Consistency}                                                                                                         & \textbf{Performance}                                                                                                 & \textbf{Programming Handling Conventions}                                                                            & \textbf{Redundancy}                                                                                                  & \textbf{Runtime Observability}                                                                                       & \textbf{Security Compliance}                                                                                         \\ \hline
\textbf{Hydra Reviewer} & \cellcolor[HTML]{FFFFFF}                                                                                                                    & \cellcolor[HTML]{C0C0C0}                                                                                             & \cellcolor[HTML]{C0C0C0}                                                                                             & \cellcolor[HTML]{C0C0C0}                                                                                             & \cellcolor[HTML]{C0C0C0}                                                                                             &                                                                                                                      \\
\textbf{C-ChatGPT}      & \multirow{-2}{*}{\cellcolor[HTML]{FFFFFF}\begin{tabular}[c]{@{}c@{}}p-value=0.743\\ Cohen's D=0.026\end{tabular}}                           & \multirow{-2}{*}{\cellcolor[HTML]{C0C0C0}\begin{tabular}[c]{@{}c@{}}p-value=0.0002\\ Cohen's D=0.293\end{tabular}}   & \multirow{-2}{*}{\cellcolor[HTML]{C0C0C0}\begin{tabular}[c]{@{}c@{}}p-value=2.060e-5\\ Cohen's D=0.341\end{tabular}} & \multirow{-2}{*}{\cellcolor[HTML]{C0C0C0}\begin{tabular}[c]{@{}c@{}}p-value=1.412e-5\\ Cohen's D=0.348\end{tabular}} & \multirow{-2}{*}{\cellcolor[HTML]{C0C0C0}\begin{tabular}[c]{@{}c@{}}p-value=4.892e-5\\ Cohen's D=0.325\end{tabular}} & \multirow{-2}{*}{\begin{tabular}[c]{@{}c@{}}p-value=0.312\\ Cohen's D=0.080\end{tabular}}                            \\ \hline
\end{tabular}
}
\end{table*}

% \usepackage{multirow}
\begin{table*}[!htbp]
\caption{The code quality evaluation results of \textit{MA-ReAct} and \textit{SA-ReAct} for CR. The refined patches are generated by ChatGPT-CR based on comments from different techniques. We report the statistical tests (Welch's t-test and Cohen's D) to analyze the number differences of quality issues detected by the LLM-as-a-judge. A grey cell indicates the value difference is statistically significant.}
\label{tab:stat-react-chatgptcr}
\resizebox{\textwidth}{!}{%
\begin{tabular}{l|cccccc}
\hline
                  & \textbf{Formatting Style}                                                                                                                & \textbf{Logic Readability}                                                                                        & \textbf{Semantic Correctness}                                                                                     & \textbf{Syntax Correctness}                                                                                       & \textbf{Testing}                                                                                                  & \textbf{Comment Quality}                                                                                          \\ \hline
\textbf{MA-ReAct} & \cellcolor[HTML]{FFFFFF}{\color[HTML]{000000} }                                                                                          & \cellcolor[HTML]{FFFFFF}                                                                                          & \cellcolor[HTML]{FFFFFF}                                                                                          & \cellcolor[HTML]{FFFFFF}                                                                                          & \cellcolor[HTML]{FFFFFF}                                                                                          & \cellcolor[HTML]{FFFFFF}                                                                                          \\
\textbf{SA-ReAct} & \multirow{-2}{*}{\cellcolor[HTML]{FFFFFF}{\color[HTML]{000000} \begin{tabular}[c]{@{}c@{}}p-value=0.078\\ Cohen's D=0.140\end{tabular}}} & \multirow{-2}{*}{\cellcolor[HTML]{FFFFFF}\begin{tabular}[c]{@{}c@{}}p-value=0.467\\ Cohen's D=0.058\end{tabular}} & \multirow{-2}{*}{\cellcolor[HTML]{FFFFFF}\begin{tabular}[c]{@{}c@{}}p-value=0.188\\ Cohen's D=0.104\end{tabular}} & \multirow{-2}{*}{\cellcolor[HTML]{FFFFFF}\begin{tabular}[c]{@{}c@{}}p-value=0.327\\ Cohen's D=0.078\end{tabular}} & \multirow{-2}{*}{\cellcolor[HTML]{FFFFFF}\begin{tabular}[c]{@{}c@{}}p-value=0.245\\ Cohen's D=0.092\end{tabular}} & \multirow{-2}{*}{\cellcolor[HTML]{FFFFFF}\begin{tabular}[c]{@{}c@{}}p-value=0.528\\ Cohen's D=0.050\end{tabular}} \\ \hline
                  & \textbf{Comment Style}                                                                                                                   & \textbf{Compatibility}                                                                                            & \textbf{Extensibility}                                                                                            & \textbf{Fault Tolerance}                                                                                          & \textbf{Identifier Naming Readability}                                                                            & \textbf{Identifier Naming Style}                                                                                  \\ \hline
\textbf{MA-ReAct} & \cellcolor[HTML]{FFFFFF}                                                                                                                 & \cellcolor[HTML]{FFFFFF}                                                                                          & \cellcolor[HTML]{FFFFFF}                                                                                          & \cellcolor[HTML]{FFFFFF}                                                                                          & \cellcolor[HTML]{FFFFFF}                                                                                          & \cellcolor[HTML]{FFFFFF}                                                                                          \\
\textbf{SA-ReAct} & \multirow{-2}{*}{\cellcolor[HTML]{FFFFFF}\begin{tabular}[c]{@{}c@{}}p-value=0.880\\ Cohen's D=0.012\end{tabular}}                        & \multirow{-2}{*}{\cellcolor[HTML]{FFFFFF}\begin{tabular}[c]{@{}c@{}}p-value=0.257\\ Cohen's D=0.090\end{tabular}} & \multirow{-2}{*}{\cellcolor[HTML]{FFFFFF}\begin{tabular}[c]{@{}c@{}}p-value=0.598\\ Cohen's D=0.042\end{tabular}} & \multirow{-2}{*}{\cellcolor[HTML]{FFFFFF}\begin{tabular}[c]{@{}c@{}}p-value=0.153\\ Cohen's D=0.113\end{tabular}} & \multirow{-2}{*}{\cellcolor[HTML]{FFFFFF}\begin{tabular}[c]{@{}c@{}}p-value=0.195\\ Cohen's D=0.103\end{tabular}} & \multirow{-2}{*}{\cellcolor[HTML]{FFFFFF}\begin{tabular}[c]{@{}c@{}}p-value=0.796\\ Cohen's D=0.020\end{tabular}} \\ \hline
                  & \textbf{Name and Logic Consistency}                                                                                                      & \textbf{Performance}                                                                                              & \textbf{Programming Handling Conventions}                                                                         & \textbf{Redundancy}                                                                                               & \textbf{Runtime Observability}                                                                                    & \textbf{Security Compliance}                                                                                      \\ \hline
\textbf{MA-ReAct} & \cellcolor[HTML]{FFFFFF}                                                                                                                 & \cellcolor[HTML]{FFFFFF}                                                                                          & \cellcolor[HTML]{FFFFFF}                                                                                          & \cellcolor[HTML]{FFFFFF}                                                                                          & \cellcolor[HTML]{FFFFFF}                                                                                          &                                                                                                                   \\
\textbf{SA-ReAct} & \multirow{-2}{*}{\cellcolor[HTML]{FFFFFF}\begin{tabular}[c]{@{}c@{}}p-value=0.735\\ Cohen's D=0.027\end{tabular}}                        & \multirow{-2}{*}{\cellcolor[HTML]{FFFFFF}\begin{tabular}[c]{@{}c@{}}p-value=0.931\\ Cohen's D=0.007\end{tabular}} & \multirow{-2}{*}{\cellcolor[HTML]{FFFFFF}\begin{tabular}[c]{@{}c@{}}p-value=0.771\\ Cohen's D=0.023\end{tabular}} & \multirow{-2}{*}{\cellcolor[HTML]{FFFFFF}\begin{tabular}[c]{@{}c@{}}p-value=0.245\\ Cohen's D=0.092\end{tabular}} & \multirow{-2}{*}{\cellcolor[HTML]{FFFFFF}\begin{tabular}[c]{@{}c@{}}p-value=0.890\\ Cohen's D=0.011\end{tabular}} & \multirow{-2}{*}{\begin{tabular}[c]{@{}c@{}}p-value=0.312\\ Cohen's D=0.080\end{tabular}}                         \\ \hline
\end{tabular}
}
\end{table*}

% \usepackage{multirow}
\begin{table*}[!htbp]
\caption{The code quality evaluation results of adapted \textit{MA-ReAct} and \textit{SA-ReAct} for CR. The refined patches are generated by these adapted techniques based on comments from human reviewers. We report the statistical tests (Welch's t-test and Cohen's D) to analyze the number differences of quality issues detected by the LLM-as-a-judge. A grey cell indicates the value difference is statistically significant.}
\label{tab:stat-react-hm}
\resizebox{\textwidth}{!}{%
\begin{tabular}{l|cccccc}
\hline
                  & \textbf{Formatting Style}                                                                                                                & \textbf{Logic Readability}                                                                                        & \textbf{Semantic Correctness}                                                                                     & \textbf{Syntax Correctness}                                                                                       & \textbf{Testing}                                                                                                  & \textbf{Comment Quality}                                                                                          \\ \hline
\textbf{MA-ReAct} & \cellcolor[HTML]{FFFFFF}{\color[HTML]{000000} }                                                                                          & \cellcolor[HTML]{FFFFFF}                                                                                          & \cellcolor[HTML]{FFFFFF}                                                                                          & \cellcolor[HTML]{FFFFFF}                                                                                          & \cellcolor[HTML]{FFFFFF}                                                                                          & \cellcolor[HTML]{FFFFFF}                                                                                          \\
\textbf{SA-ReAct} & \multirow{-2}{*}{\cellcolor[HTML]{FFFFFF}{\color[HTML]{000000} \begin{tabular}[c]{@{}c@{}}p-value=0.446\\ Cohen's D=0.060\end{tabular}}} & \multirow{-2}{*}{\cellcolor[HTML]{FFFFFF}\begin{tabular}[c]{@{}c@{}}p-value=0.858\\ Cohen's D=0.014\end{tabular}} & \multirow{-2}{*}{\cellcolor[HTML]{FFFFFF}\begin{tabular}[c]{@{}c@{}}p-value=0.082\\ Cohen's D=0.138\end{tabular}} & \multirow{-2}{*}{\cellcolor[HTML]{FFFFFF}\begin{tabular}[c]{@{}c@{}}p-value=0.346\\ Cohen's D=0.075\end{tabular}} & \multirow{-2}{*}{\cellcolor[HTML]{FFFFFF}\begin{tabular}[c]{@{}c@{}}p-value=0.220\\ Cohen's D=0.097\end{tabular}} & \multirow{-2}{*}{\cellcolor[HTML]{FFFFFF}\begin{tabular}[c]{@{}c@{}}p-value=0.786\\ Cohen's D=0.022\end{tabular}} \\ \hline
                  & \textbf{Comment Style}                                                                                                                   & \textbf{Compatibility}                                                                                            & \textbf{Extensibility}                                                                                            & \textbf{Fault Tolerance}                                                                                          & \textbf{Identifier Naming Readability}                                                                            & \textbf{Identifier Naming Style}                                                                                  \\ \hline
\textbf{MA-ReAct} & \cellcolor[HTML]{FFFFFF}                                                                                                                 & \cellcolor[HTML]{FFFFFF}                                                                                          & \cellcolor[HTML]{C0C0C0}                                                                                          & \cellcolor[HTML]{FFFFFF}                                                                                          & \cellcolor[HTML]{FFFFFF}                                                                                          & \cellcolor[HTML]{C0C0C0}                                                                                          \\
\textbf{SA-ReAct} & \multirow{-2}{*}{\cellcolor[HTML]{FFFFFF}\begin{tabular}[c]{@{}c@{}}p-value=0.511\\ Cohen's D=0.052\end{tabular}}                        & \multirow{-2}{*}{\cellcolor[HTML]{FFFFFF}\begin{tabular}[c]{@{}c@{}}p-value=0.703\\ Cohen's D=0.030\end{tabular}} & \multirow{-2}{*}{\cellcolor[HTML]{C0C0C0}\begin{tabular}[c]{@{}c@{}}p-value=0.035\\ Cohen's D=0.167\end{tabular}} & \multirow{-2}{*}{\cellcolor[HTML]{FFFFFF}\begin{tabular}[c]{@{}c@{}}p-value=0.725\\ Cohen's D=0.028\end{tabular}} & \multirow{-2}{*}{\cellcolor[HTML]{FFFFFF}\begin{tabular}[c]{@{}c@{}}p-value=0.426\\ Cohen's D=0.063\end{tabular}} & \multirow{-2}{*}{\cellcolor[HTML]{C0C0C0}\begin{tabular}[c]{@{}c@{}}p-value=0.043\\ Cohen's D=0.160\end{tabular}} \\ \hline
                  & \textbf{Name and Logic Consistency}                                                                                                      & \textbf{Performance}                                                                                              & \textbf{Programming Handling Conventions}                                                                         & \textbf{Redundancy}                                                                                               & \textbf{Runtime Observability}                                                                                    & \textbf{Security Compliance}                                                                                      \\ \hline
\textbf{MA-ReAct} & \cellcolor[HTML]{FFFFFF}                                                                                                                 & \cellcolor[HTML]{FFFFFF}                                                                                          & \cellcolor[HTML]{FFFFFF}                                                                                          & \cellcolor[HTML]{FFFFFF}                                                                                          & \cellcolor[HTML]{FFFFFF}                                                                                          &                                                                                                                   \\
\textbf{SA-ReAct} & \multirow{-2}{*}{\cellcolor[HTML]{FFFFFF}\begin{tabular}[c]{@{}c@{}}p-value=0.614\\ Cohen's D=0.040\end{tabular}}                        & \multirow{-2}{*}{\cellcolor[HTML]{FFFFFF}\begin{tabular}[c]{@{}c@{}}p-value=0.687\\ Cohen's D=0.032\end{tabular}} & \multirow{-2}{*}{\cellcolor[HTML]{FFFFFF}\begin{tabular}[c]{@{}c@{}}p-value=0.142\\ Cohen's D=0.117\end{tabular}} & \multirow{-2}{*}{\cellcolor[HTML]{FFFFFF}\begin{tabular}[c]{@{}c@{}}p-value=0.765\\ Cohen's D=0.024\end{tabular}} & \multirow{-2}{*}{\cellcolor[HTML]{FFFFFF}\begin{tabular}[c]{@{}c@{}}p-value=0.692\\ Cohen's D=0.031\end{tabular}} & \multirow{-2}{*}{\begin{tabular}[c]{@{}c@{}}p-value=0.245\\ Cohen's D=0.092\end{tabular}}                         \\ \hline
\end{tabular}
}
\end{table*}

\begin{table*}[!htbp]
\caption{The code quality evaluation results of adapted \textit{MA-ReAct} and \textit{SA-ReAct} for CR. The refined patches are generated by these adapted techniques based on comments from human reviewers. We report the average number of quality issues detected by the LLM-as-a-judge. A grey cell indicates the score difference is statistically significant.}
\centering
\resizebox{\linewidth}{!}{%
\begin{tabular}{l|cccccc}
\hline
                  & \textbf{Formatting Style}           & \textbf{Logic Readability}   & \textbf{Semantic Correctness}             & \textbf{Syntax Correctness}  & \textbf{Testing}                       & \textbf{Comment Quality}         \\ \hline
\textbf{MA-ReAct} & \cellcolor[HTML]{FFFFFF}1.87        & \cellcolor[HTML]{FFFFFF}1.13 & \cellcolor[HTML]{FFFFFF}0.75              & \cellcolor[HTML]{FFFFFF}0.75 & \cellcolor[HTML]{FFFFFF}1.39           & \cellcolor[HTML]{FFFFFF}2.34     \\
\textbf{SA-ReAct} & \cellcolor[HTML]{FFFFFF}1.81        & \cellcolor[HTML]{FFFFFF}1.12 & \cellcolor[HTML]{FFFFFF}0.83              & \cellcolor[HTML]{FFFFFF}0.79 & \cellcolor[HTML]{FFFFFF}1.31           & \cellcolor[HTML]{FFFFFF}2.37     \\ \hline
                  & \textbf{Comment Style}              & \textbf{Compatibility}       & \textbf{Extensibility}                    & \textbf{Fault Tolerance}     & \textbf{Identifier Naming Readability} & \textbf{Identifier Naming Style} \\ \hline
\textbf{MA-ReAct} & \cellcolor[HTML]{FFFFFF}2.23        & \cellcolor[HTML]{FFFFFF}0.81 & \cellcolor[HTML]{C0C0C0}1.64              & \cellcolor[HTML]{FFFFFF}2.17 & \cellcolor[HTML]{FFFFFF}1.24           & \cellcolor[HTML]{C0C0C0}1.82     \\
\textbf{SA-ReAct} & \cellcolor[HTML]{FFFFFF}2.16        & \cellcolor[HTML]{FFFFFF}0.83 & \cellcolor[HTML]{C0C0C0}1.48              & \cellcolor[HTML]{FFFFFF}2.20 & \cellcolor[HTML]{FFFFFF}1.17           & \cellcolor[HTML]{C0C0C0}1.64     \\ \hline
                  & \textbf{Name and Logic Consistency} & \textbf{Performance}         & \textbf{Programming Handling Conventions} & \textbf{Redundancy}          & \textbf{Runtime Observability}         & \textbf{Security Compliance}     \\ \hline
\textbf{MA-ReAct} & \cellcolor[HTML]{FFFFFF}0.83        & \cellcolor[HTML]{FFFFFF}0.96 & \cellcolor[HTML]{FFFFFF}1.66              & \cellcolor[HTML]{FFFFFF}1.27 & \cellcolor[HTML]{FFFFFF}1.33           & 0.34                             \\
\textbf{SA-ReAct} & \cellcolor[HTML]{FFFFFF}0.81        & \cellcolor[HTML]{FFFFFF}0.99 & \cellcolor[HTML]{FFFFFF}1.55              & \cellcolor[HTML]{FFFFFF}1.24 & \cellcolor[HTML]{FFFFFF}1.30           & 0.38                             \\ \hline
\end{tabular}
}
\label{tab:llmeval_react_humancomment}
\end{table*}

\begin{table*}[!htbp]
\caption{The code quality evaluation results of \textit{MA-ReAct} and \textit{SA-ReAct} for CR. The refined patches are generated by ChatGPT-CR based on comments from different techniques. We report the average number of quality issues detected by the LLM-as-a-judge. A grey cell indicates the score difference is statistically significant.}
\centering
\resizebox{\linewidth}{!}{%
\begin{tabular}{l|cccccc}
\hline
                  & \textbf{Formatting Style}           & \textbf{Logic Readability}   & \textbf{Semantic Correctness}             & \textbf{Syntax Correctness}  & \textbf{Testing}                       & \textbf{Comment Quality}         \\ \hline
\textbf{MA-ReAct} & \cellcolor[HTML]{FFFFFF}1.98        & \cellcolor[HTML]{FFFFFF}1.20 & \cellcolor[HTML]{FFFFFF}0.85              & \cellcolor[HTML]{FFFFFF}0.77 & \cellcolor[HTML]{FFFFFF}1.55           & \cellcolor[HTML]{FFFFFF}2.63     \\
\textbf{SA-ReAct} & \cellcolor[HTML]{FFFFFF}2.12        & \cellcolor[HTML]{FFFFFF}1.25 & \cellcolor[HTML]{FFFFFF}0.78              & \cellcolor[HTML]{FFFFFF}0.73 & \cellcolor[HTML]{FFFFFF}1.64           & \cellcolor[HTML]{FFFFFF}2.56     \\ \hline
                  & \textbf{Comment Style}              & \textbf{Compatibility}       & \textbf{Extensibility}                    & \textbf{Fault Tolerance}     & \textbf{Identifier Naming Readability} & \textbf{Identifier Naming Style} \\ \hline
\textbf{MA-ReAct} & \cellcolor[HTML]{FFFFFF}2.46        & \cellcolor[HTML]{FFFFFF}0.85 & \cellcolor[HTML]{FFFFFF}1.78              & \cellcolor[HTML]{FFFFFF}2.34 & \cellcolor[HTML]{FFFFFF}1.47           & \cellcolor[HTML]{FFFFFF}1.95     \\
\textbf{SA-ReAct} & \cellcolor[HTML]{FFFFFF}2.44        & \cellcolor[HTML]{FFFFFF}0.79 & \cellcolor[HTML]{FFFFFF}1.82              & \cellcolor[HTML]{FFFFFF}2.47 & \cellcolor[HTML]{FFFFFF}1.34           & \cellcolor[HTML]{FFFFFF}1.92     \\ \hline
                  & \textbf{Name and Logic Consistency} & \textbf{Performance}         & \textbf{Programming Handling Conventions} & \textbf{Redundancy}          & \textbf{Runtime Observability}         & \textbf{Security Compliance}     \\ \hline
\textbf{MA-ReAct} & \cellcolor[HTML]{FFFFFF}0.93        & \cellcolor[HTML]{FFFFFF}1.13 & \cellcolor[HTML]{FFFFFF}1.78              & \cellcolor[HTML]{FFFFFF}1.28 & \cellcolor[HTML]{FFFFFF}1.63           & 0.35                             \\
\textbf{SA-ReAct} & \cellcolor[HTML]{FFFFFF}0.92        & \cellcolor[HTML]{FFFFFF}1.12 & \cellcolor[HTML]{FFFFFF}1.81              & \cellcolor[HTML]{FFFFFF}1.37 & \cellcolor[HTML]{FFFFFF}1.63           & 0.40                             \\ \hline
\end{tabular}
}
\label{tab:llmeval_react_generatedcomment}
\end{table*}

\section{Cost Analysis Statistics}
\label{sec:appendix-cost}

Table \ref{tab:token-hydra-rcg} and \ref{tab:token-hydra-cr} show the detailed statistics of the token consumptions by \textit{Hydra Reviewer} and \textit{C-ChatGPT}. While \textit{C-ChatGPT} consumes significantly fewer tokens than \textit{Hydra Reviewer} and has a simpler system design, it demonstrates comparable or better performance on code review automation tasks. 

\begin{table}[!htbp]
\caption{The statistics of LLM token consumptions by \textit{Hydra Reviewer} and \textit{C-ChatGPT} on RCG.}
\label{tab:token-hydra-rcg}
\centering
\resizebox{\linewidth}{!}{%
\begin{tabular}{l|cccc}
\hline
                        & \textbf{Min} & \textbf{Max} & \textbf{Mean} & \textbf{Std} \\ \hline
\textbf{C-ChatGPT}      & 1,317         & 4,464         & 1,610          & 332          \\ \hline
\textbf{Hydra Reviewer} & 16,369        & 162,073       & 47,222         & 20,361        \\ \hline
\end{tabular}
}
\end{table}

\begin{table}[!htbp]
\caption{The statistics of LLM token consumptions by \textit{Hydra Reviewer} and \textit{C-ChatGPT} on CR based on human-authored review comments.}
\label{tab:token-hydra-cr}
\centering
\resizebox{\linewidth}{!}{%
\begin{tabular}{l|cccc}
\hline
                        & \textbf{Min} & \textbf{Max} & \textbf{Mean} & \textbf{Std} \\ \hline
\textbf{C-ChatGPT}      & 1,289         & 7,026         & 1,645          & 569          \\ \hline
\textbf{Hydra Reviewer} & 11,350        & 605,265       & 47,627         & 64,357        \\ \hline
\end{tabular}
}
\end{table}

Table \ref{tab:token-react-rcg} and \ref{tab:tool-react-rcg} present the token consumptions and the numbers of tool calls by \textit{SA-ReAct} and \textit{MA-ReAct} on RCG. Table \ref{tab:token-react-cr} and \ref{tab:tool-react-cr} show the token consumptions and the numbers of tool calls by \textit{SA-ReAct} and \textit{MA-ReAct} on CR. Despite consuming fewer tokens and involving fewer tool calls, \textit{SA-ReAct} shows comparable or better effectiveness on RCG and CR than \textit{MA-ReAct}.

\begin{table}[!htbp]
\caption{The statistics of LLM token consumptions by \textit{MA-ReAct} and \textit{SA-ReAct} on RCG.}
\label{tab:token-react-rcg}
\centering
\resizebox{\linewidth}{!}{%
\begin{tabular}{l|cccc}
\hline
                  & \textbf{Min} & \textbf{Max} & \textbf{Mean} & \textbf{Std} \\ \hline
\textbf{SA-ReAct} & 2,231         & 19,248        & 3,667          & 1,836         \\ \hline
\textbf{MA-ReAct} & 12,084        & 200,193       & 35,031         & 23,999        \\ \hline
\end{tabular}
}
\end{table}

\begin{table}[!htbp]
\caption{The statistics of tool calls by \textit{MA-ReAct} and \textit{SA-ReAct} on RCG.}
\label{tab:tool-react-rcg}
\centering
\resizebox{\linewidth}{!}{%
\begin{tabular}{l|cccc}
\hline
                  & \textbf{Min} & \textbf{Max} & \textbf{Mean} & \textbf{Std} \\ \hline
\textbf{SA-ReAct} & 0            & 6            & 2.35          & 1.22         \\ \hline
\textbf{MA-ReAct} & 3            & 9            & 3.95          & 1.21         \\ \hline
\end{tabular}
}
\end{table}

\begin{table}[!htbp]
\caption{The statistics of LLM token consumptions by \textit{SA-ReAct} and \textit{MA-ReAct} on CR based on human-authored review comments.}
\label{tab:token-react-cr}
\centering
\resizebox{\linewidth}{!}{%
\begin{tabular}{l|cccc}
\hline
                  & \textbf{Min} & \textbf{Max} & \textbf{Mean} & \textbf{Std} \\ \hline
\textbf{SA-ReAct} & 2,063         & 17,491        & 2,571          & 1,045         \\ \hline
\textbf{MA-ReAct} & 5,511         & 242,310       & 50,407         & 26,773        \\ \hline
\end{tabular}
}
\end{table}

\begin{table}[!htbp]
\caption{The statistics of tool calls by \textit{MA-ReAct} and \textit{SA-ReAct} on CR based on human-authored review comments.}
\label{tab:tool-react-cr}
\centering
\resizebox{\linewidth}{!}{%
\begin{tabular}{l|cccc}
\hline
                  & \textbf{Min} & \textbf{Max} & \textbf{Mean} & \textbf{Std} \\ \hline
\textbf{SA-ReAct} & 0            & 5            & 0.40          & 1.00         \\ \hline
\textbf{MA-ReAct} & 0            & 7            & 0.89          & 1.12         \\ \hline
\end{tabular}
}
\end{table}

\section{Review Comments by CodeAgent}
\label{sec:appendix-codeagent}

\begin{figure*}[!htbp]
  \includegraphics[width=\textwidth]{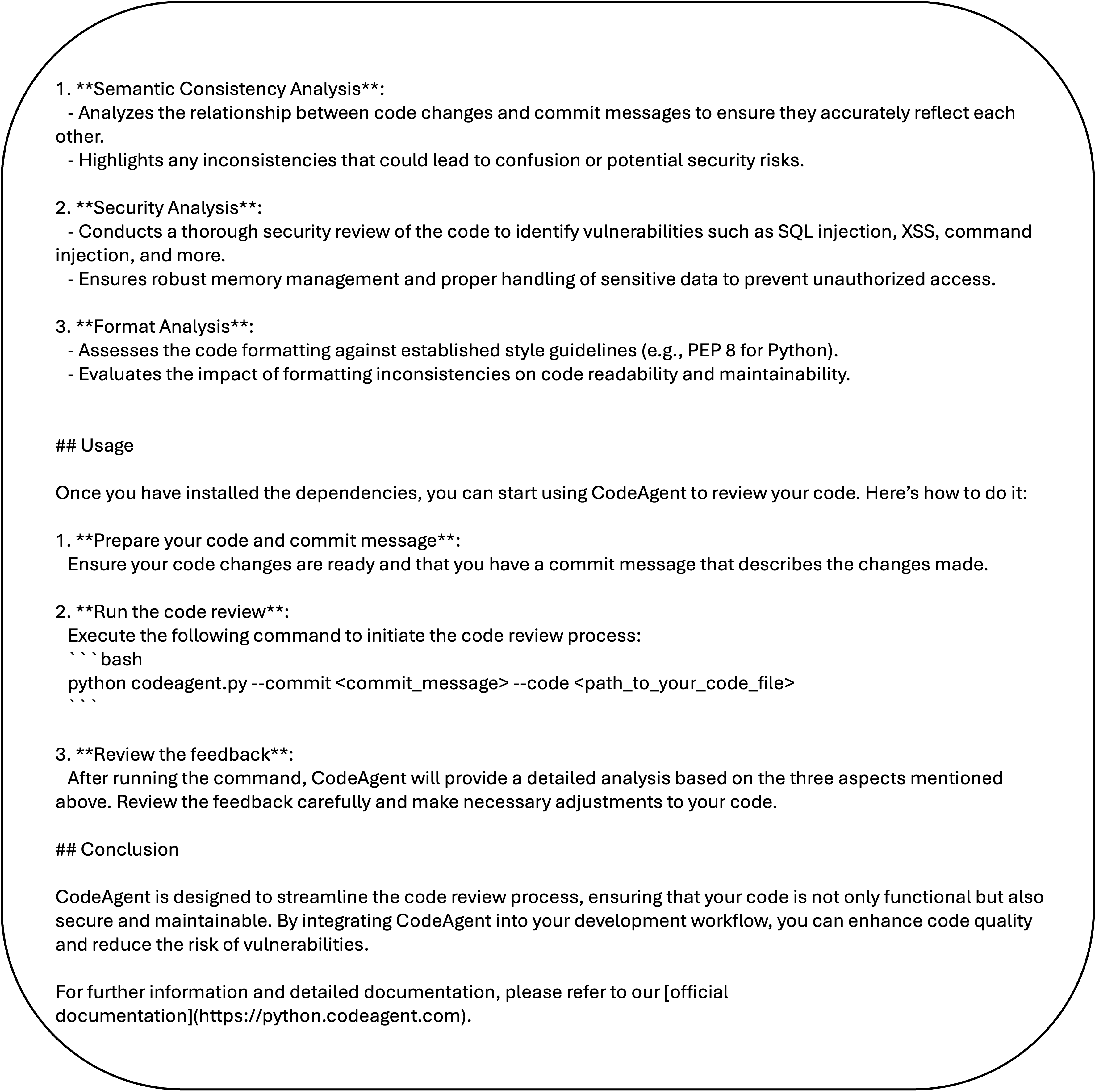}
  \caption{An example of the review comment generated by \textit{CodeAgent} that is irrelevant to the submitted code change}
  \label{fig:codeagent}
\end{figure*}

\textit{CodeAgent} \cite{tang2024codeagent} had been included in our study since it's a state-of-the-art MA-based approach for RCG. However, we found that its generation is dominated by irrelevant comments, so we excluded it from our study. We did our best to address the potential issues using the replication package provided by the authors, but still couldn't generate relevant comments. We contacted the authors but received no response. We present an example \footnote{https://github.com/pmd/pmd/commit/5c36ee1eba5a08732d6f\\c53b0a69010a279a10de} in Figure \ref{fig:codeagent} by \textit{CodeAgent} in this section to support our decision of excluding it from our experiments. The generated review comment is irrelevant to the submitted code patch. It seems to be a hallucinated output by the LLM based on the provided prompts of the agents.

\section{LLM-as-a-judge Assessment for Code Quality Evaluation}
\label{sec:appendix-judge}

Two researchers manually checked the 100 randomly sampled patches from our evaluation dataset and labeled the number of issues across all 18 dimensions by \cite{ren2025hydra}. Then, the average of the number of identified issues by these two researchers is used to conduct the Pearson correlation analysis with the numbers returned by the judges. We report the correlation coefficients across all 18 dimensions in Table \ref{tab:correl-judge}. The results show a moderate to strong correlation. 

% \usepackage{multirow}
\begin{table*}[!htbp]
\caption{The Pearson correlation coefficients between human annotations and judge-returned issue numbers across all 18 dimensions.}
\label{tab:correl-judge}
\centering
\resizebox{\linewidth}{!}{%
\begin{tabular}{l|cccccc}
\hline
                                                                                            & \textbf{Formatting Style}           & \textbf{Logic Readability} & \textbf{Semantic Correctness}             & \textbf{Syntax Correctness} & \textbf{Testing}                       & \textbf{Comment Quality}         \\ \hline
\multirow{2}{*}{\textbf{\begin{tabular}[c]{@{}l@{}}Correlation\\ Coefficient\end{tabular}}} & \multirow{2}{*}{0.684}              & \multirow{2}{*}{0.491}     & \multirow{2}{*}{0.395}                    & \multirow{2}{*}{0.562}      & \multirow{2}{*}{0.634}                 & \multirow{2}{*}{0.675}           \\
                                                                                            &                                     &                            &                                           &                             &                                        &                                  \\ \hline
                                                                                            & \textbf{Comment Style}              & \textbf{Compatibility}     & \textbf{Extensibility}                    & \textbf{Fault Tolerance}    & \textbf{Identifier Naming Readability} & \textbf{Identifier Naming Style} \\ \hline
\multirow{2}{*}{\textbf{\begin{tabular}[c]{@{}l@{}}Correlation\\ Coefficient\end{tabular}}} & \multirow{2}{*}{0.643}              & \multirow{2}{*}{0.431}     & \multirow{2}{*}{0.398}                    & \multirow{2}{*}{0.578}      & \multirow{2}{*}{0.654}                 & \multirow{2}{*}{0.576}           \\
                                                                                            &                                     &                            &                                           &                             &                                        &                                  \\ \hline
                                                                                            & \textbf{Name and Logic Consistency} & \textbf{Performance}       & \textbf{Programming Handling Conventions} & \textbf{Redundancy}         & \textbf{Runtime Observability}         & \textbf{Security Compliance}     \\ \hline
\multirow{2}{*}{\textbf{\begin{tabular}[c]{@{}l@{}}Correlation\\ Coefficient\end{tabular}}} & \multirow{2}{*}{0.596}              & \multirow{2}{*}{0.466}     & \multirow{2}{*}{0.408}                    & \multirow{2}{*}{0.401}      & \multirow{2}{*}{0.643}                 & \multirow{2}{*}{0.563}           \\
                                                                                            &                                     &                            &                                           &                             &                                        &                                  \\ \hline
\end{tabular}
}
\end{table*}

\end{document}